\documentclass[10pt,journal,compsoc]{IEEEtran}

\usepackage{booktabs} 
\usepackage{xcolor,colortbl}
\definecolor{Gray}{gray}{0.9}
\definecolor{Green}{RGB}{89,163,76}
\usepackage{fontawesome}
\usepackage[caption=false,font=footnotesize,labelfont=sf,textfont=sf]{subfig}
\usepackage{url}
\usepackage{todonotes}
\usepackage{balance}
\usepackage{framed}
\usepackage{textcomp}
\usepackage{amssymb}
\usepackage{microtype}
\usepackage{xfrac}

\usepackage[nocompress]{cite}

\newcommand{\rqn}[1]{\textbf{RQ{#1}}}
\newcommand{\rquestion}[2]{\begin{framed}%
  \noindent\emph{\rqn{#1}\textbf{:} #2}
\end{framed}}

\newcommand{\highlight}[1]{\begin{framed}%
  \noindent\emph{#1}
\end{framed}}

\renewcommand{\today}{September 10, 2019}
\renewcommand{\check}{{\color{Green}$\checkmark$}}
\newcommand{\no}{{\color{red}$\times$}}
\newcommand{\na}{n.a.}
\newcommand{\on}{{\color{Green}\faToggleOn}}
\newcommand{\off}{{\color{gray}\faToggleOff}}

\begin{document}

\title{On the Energy Footprint of Mobile Testing Frameworks}

\author{Lu\'{i}s~Cruz,~\IEEEmembership{Member,~IEEE,}
        and~Rui~Abreu,~\IEEEmembership{Senior~Member,~IEEE,}

\IEEEcompsocitemizethanks{\IEEEcompsocthanksitem Lu\'{i}s Cruz is with Delft University of Technology.\protect\\
E-mail: see https://luiscruz.github.io/
\IEEEcompsocthanksitem  Rui Abreu is with University of Lisbon and INESC-ID.\protect\\
E-mail: see https://ruimaranhao.com/}
\thanks{Manuscript received April 19, 2005; revised August 26, 2015.}}

%
\markboth{IEEE Transactions on Software Engineering (TSE),~Vol.~XX, No.~X, August~20XX}%
{Cruz L. and Abreu R.: On the Energy Footprint of Mobile Testing Frameworks}

\IEEEtitleabstractindextext{%
\begin{abstract}

High energy consumption is a challenging issue that an ever increasing
number of mobile applications face today.
However, energy consumption is being tested in an \emph{ad hoc} way,
 despite being an important non-functional requirement of
an application.
Such limitation becomes
particularly disconcerting during software testing: on the one hand, developers
do not really know how to measure energy; on the other hand, there is no
knowledge as to what is the energy overhead imposed by the testing framework.
In this paper, as we evaluate eight popular mobile UI automation frameworks, we
have discovered that there are automation frameworks that increase energy
consumption up to roughly 2200\%. While limited in the interactions one can do,
\emph{Espresso} is the most energy efficient framework. However, depending on
the needs of the tester, \emph{Appium}, \emph{Monkeyrunner}, or
\emph{UIAutomator} are good alternatives. In practice, results show that deciding
which is the most suitable framework is vital. We provide a decision tree to
help developers make an educated decision on which framework suits best their
testing needs.

\end{abstract}

}
\maketitle
\IEEEdisplaynontitleabstractindextext
\IEEEpeerreviewmaketitle

\IEEEraisesectionheading{\section{Introduction}\label{sec:intro}}

The popularity of mobile applications (also known as \textit{apps}) has brought a unique,
non-functional concern to the attention of developers -- energy efficiency~\cite{ferreira2013revisiting}. Mobile apps
that (quickly) drain the battery of mobile devices are perceived as being of poor
quality by users\footnote{\emph{Apigee}'s survey on the reasons leading to bad mobile
app reviews: \url{https://cloud.google.com/apigee/news} (visited on \today{})}. As a
consequence, users are likely to uninstall an app even if it provides useful
functionality and there is no better alternative. In fact, mobile network
operators recommend users to uninstall apps that are energy
inefficient\footnote{\emph{High Risk Android Apps}:
\url{https://www.verizonwireless.com/support/services-and-apps/} (visited on \today{})}.
It is therefore important to provide developers with tools and knowledge to
ship energy efficient mobile apps~\cite{pang2016programmers,pereira2017helping,linares2017continuous,cruz2019catalog}.



Automated testing tools help validate not
only functional but also non-functional requirements such as
scalability and usability~\cite{morgado2015impact,moreira2013pattern}.
When it comes to energy testing, the most reliable approach to measure
the energy consumption of mobile software is by using user interface (UI) automation
frameworks~\cite{li2014empirical,lee2015entrack,linares2014mining,di2017software,carette2017investigating,cao2017deconstructing}.
These frameworks are used to mimic user interaction in mobile apps while using an energy profiling tool.
An alternative is to use manual testing but it creates bias, is error prone, and is
both time and human resource consuming~\cite{rasmussen2014green}.

While using a UI automation framework is the most
suitable option to test apps, there are still energy-related concerns that need
to be addressed. By replicating interactions, frameworks are
bypassing or creating overhead on system behavior. For instance, before
executing a \emph{Tap}\footnote{\emph{Tap} is a gesture in which a user taps the screen with his
finger.}, it is necessary to programmatically look up the target UI component. This
creates extra processing that would not happen during an ordinary execution of
the app. These idiosyncrasies are addressed in the work proposed in this paper,
as they may have a negative impact on energy consumption results.

As a motivational example, consider the following scenario: an app provides a
\emph{tweet} feed that consists of a list of \emph{tweets} including their
media content (such as, pictures, GIFs, videos, URLs). The product owner
noticed that users rather value apps with low energy consumption. Hence,
the development team has to address this non-functional requirement.

\begin{figure}
  \centering
\includegraphics[width=0.75\linewidth]{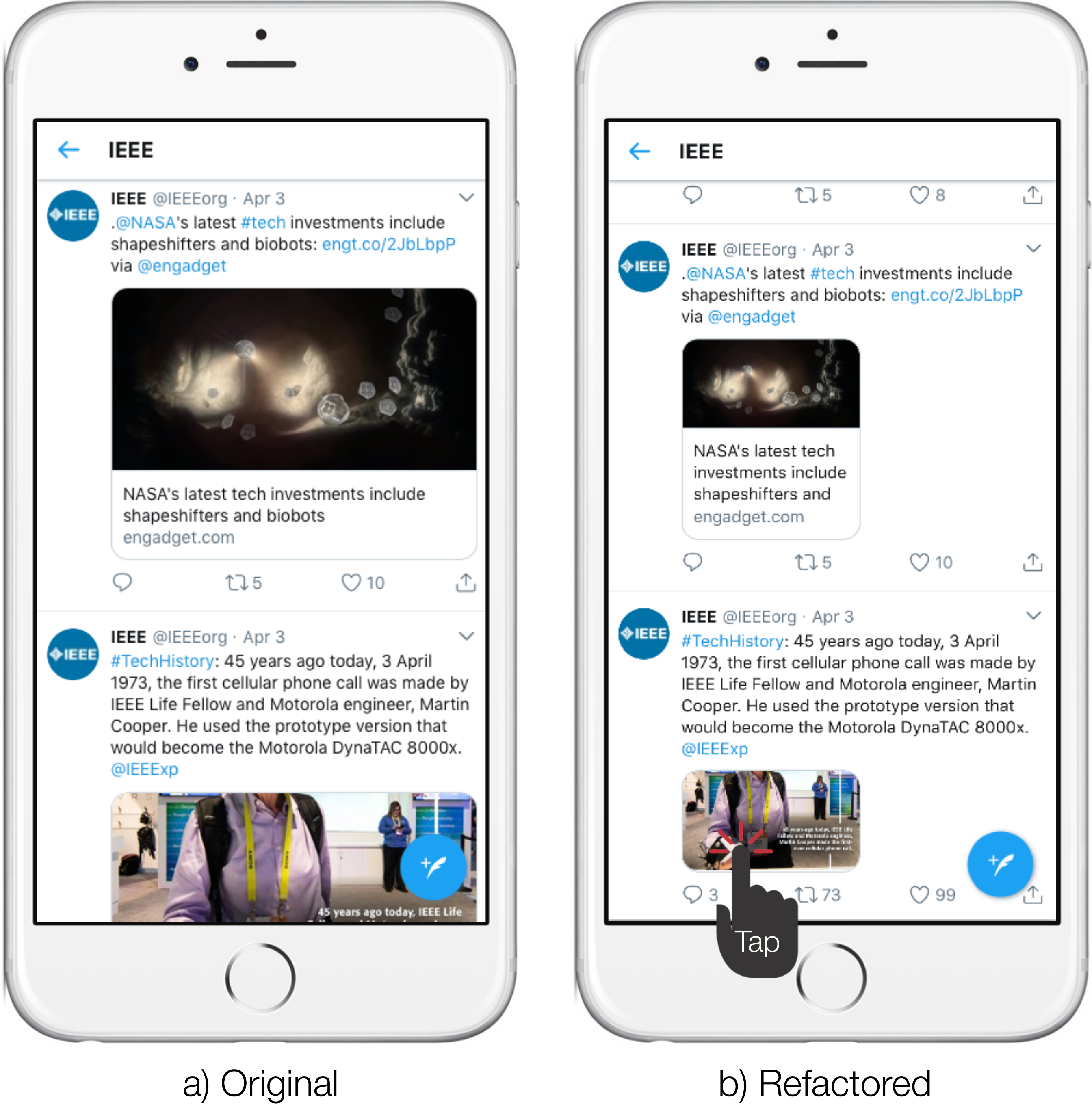}
\caption{Two versions of the example app.}
\label{fig:example_app}
\end{figure}

One idea is to show plain text and pictures with low resolution. Original media
content would be rendered upon a user \emph{Tap} on the \emph{tweet}, as
depicted in Figure~\ref{fig:example_app}. With this approach, energy is
potentially saved by rendering only media that the user is interested in. To
validate this solution, developers created automated scripts to
mimic user interaction in both versions of the app while measuring the energy
consumption using a power meter. The script for the original version consisted
in opening the app and scroll the next 20 items, whereas the new version's
script consisted in opening the app and scrolling the next 20 items while
tapping in 5 of them (a number they agreed to be the average hit rate of their
users). A problem that arises is that the automation framework spends more
energy to perform the five extra \emph{Taps}. Imagining that for each
\emph{Tap} the automation framework consumes $1$ joule\footnote{Joule (J) is
the energy unit in the International System of Units.} (J), the new version
will have to spend at least $5\textup{J}$ less than the original version to be
perceived as more efficient. If not, it gets rejected even though the new
version could be more efficient.

More efficient frameworks could reduce this threshold to a more insignificant
value. However, since automation frameworks have not considered energy
consumption as an issue, developers do not have a sense of which framework is
more suitable to perform reliable energy measurements.

The primary goal of this work is to study popular UI automation frameworks in the
context of testing the energy efficiency of mobile apps. We address the
following research questions:

\begin{itemize}

\item \rqn{1}: Does the energy consumption overhead created by UI automation
frameworks affect the results of the energy efficiency of mobile applications?

\item \rqn{2}: What is the most suitable framework to profile energy consumption?

\item \rqn{3}: Are there any best practices when it comes to creating automated scripts
for energy efficiency tests?

\end{itemize}

We measure the energy consumption of common user interactions: \emph{Tap},
\emph{Long Tap}, \emph{Drag And Drop}, \emph{Swipe}, \emph{Pinch \& Spread},
\emph{Back button}, \emph{Input text}, \emph{Find by id}, \emph{Find by description}, and
\emph{Find by content}.

Results show that \emph{Espresso} is the framework with best
energy footprint, although \emph{Appium}, \emph{Monkeyrunner}, and
\emph{UIAutomator} are also good candidates. On the other side of the spectrum
are \emph{AndroidViewClient} and \emph{Calabash}, which makes them not suitable
to test the energy efficiency of apps yet. For a general purpose context,
\emph{Appium} follows as being the best candidate. We have further discovered
that methods that use content to look up UI components need to be avoided since
they are not energy savvy.

\highlight{
Overheads incurred by UI automation frameworks ought to be
considered when measuring energy consumption of mobile apps.
}

To sum up, the main contributions of this paper are:
\begin{itemize}

  \item A comprehensive study on energy consumption of user interactions
  mimicked by UI automation frameworks.

  \item Comparison of the state-of-the-art UI automation frameworks and their
  features in the context of energy tests.

  \item Best practices regarding the API usage of the framework for energy tests, including a decision tree to
  help choose the framework which suits one needs.

\end{itemize}

\section{Related Work}


UI automation frameworks play an important role on the research of mobile software
energy efficiency. They are used as part of the experimental setup for the
validation of approaches for energy efficiency of mobile apps.
\emph{Monkeyrunner} has been used to assess the energy efficiency of Android's API
usage patterns~\cite{linares2014mining}. It was found that UI manipulation
tasks (e.g., method \texttt{findViewById}) and database operations are expensive in
terms of energy consumption. These findings suggest that UI automation
frameworks might as well create a considerable overhead on energy consumption.
\emph{Monkeyrunner} has also been used to assess benefits in energy efficiency
on the usage of Progressive Web Apps technology in mobile web
apps~\cite{malavolta2017assessing}, despite the fact that no statistically
significant differences were found. \emph{Android View Client} has
been used to assess energy efficiency improvements of performance based
optimizations for Android applications~\cite{cruz2017performance,cruz2018using}, being able
to improve energy consumption up to $5$\% in real, mature Android applications.
Other works have also used \emph{Robotium}~\cite{hecht2016empirical},
\emph{Calabash}~\cite{cao2017deconstructing,carette2017investigating}, and
\emph{RERAN}~\cite{gomez2013reran,sahin2016benchmarks}. Our work uses a similar
approach for assessing and validating energy efficiency, but it
has distinct goals as we focus on the impact of UI automation
frameworks on energy efficiency results.

Previous work studied five Android testing frameworks in terms of fragilities
induced by maintainability~\cite{coppola2016automated,coppola2017fragility}.
Five possible threats that could break tests were identified: 1) identifier
change, 2) text change, 3) deletion or relocation of UI elements, 4) deprecated
use of physical buttons, and 5) graphics change (mainly for image recognition
testing techniques). These threats are aligned with efforts from existing
works~\cite{gao2016sitar}. Our paper differentiates itself by focusing on the
energy efficiency of Android testing tools.

In a study comparing \emph{Appium}, \emph{MonkeyTalk}, \emph{Ranorex},
\emph{Robotium}, and \emph{UIAutomator},
\emph{Robotium} and \emph{MonkeyTalk} stood out as being the best frameworks
for being easy to learn and providing a
more efficient comparison output between expected and actual
result~\cite{gunasekaran2015survey}. A similar approach was taken in other
works~\cite{kulkarni2016deployment,liu2017mechanism} but although they provide
useful insights about architecture and feature set, no
systematic comparison was conducted. We compare different frameworks with a
quantitative approach to prevent bias of results.

Linares-V{\'a}squez M. \emph{et al.} (2017) have studied the current state-of-the-art in
terms of the frameworks, tools, and services available to aid developers in
mobile testing~\cite{linares2017continuous}. It focused on 1) Automation
APIs/Frameworks, 2) Record and Replay Tools, 3) Automated Test Input Generation
Techniques, 4) Bug and Error Reporting/Monitoring Tools, 5) Mobile Testing
Services, and 6) Device Streaming Tools. It envisions that automated testing
tools mobile apps should address development restrictions: 1) restricted
time/budget for testing, 2) needs for diverse types of testing (e.g., energy),
and 3) pressure from users for continuous delivery. In a similar work, these
issues were addressed by surveying 102 developers of Android open source
projects~\cite{linares2017developers}. This work identified a need for
automatically generated test cases that can be easily maintained
over time, low-overhead tools that can be integrated with the
development workflow, and expressive test cases. Our work differs from
these studies by providing an empirical comparison solely on UI automation frameworks,
and addressing energy tests.

Choudhary R., \emph{et al.} (2015) compared automated input generation
(AIG) techniques using four metrics~\cite{choudhary2015automated}: ease
of use, ability to work on multiple platforms, code coverage, and ability to
detect faults. It was found that random exploration strategies by
\emph{Monkey}\footnote{\emph{UI/Application Exerciser Monkey} also known as \texttt{Monkey} tool: \url{https://developer.android.com/studio/test/monkey.html} (visited on \today{}).} or
\emph{Dynodroid}~\cite{machiry2013dynodroid} were more effective than more
sophisticated approaches. Although our work does not scope AIG tools, very
often they use UI automation frameworks (e.g., \emph{UIAutomator} and
\emph{Robotium}) underneath their
systems~\cite{linares2015enabling,hao2014puma,mahmood2014evodroid,liu2014capture}.
Results and insights about energy consumption in our study may also apply
to tools that build on top of UI automation frameworks.

\section{Design of the Empirical Study}
\label{sec:methodology}

To answer the research questions outlined in the Introduction, we designed an
experimental setup to automatically measure energy consumption of Android apps.
In particular, our methodology consists in the following steps:

\begin{enumerate}
  \item Preparation of an Android device to use with a power monitor.
  \item Creation of a stack of UI interaction scripts for all frameworks.
  \item Automation of the execution of tests for each framework to run in batch mode.
  \item Collection and analysis of data.
\end{enumerate}

\begin{figure}
  \centering
\includegraphics[width=0.8\linewidth]{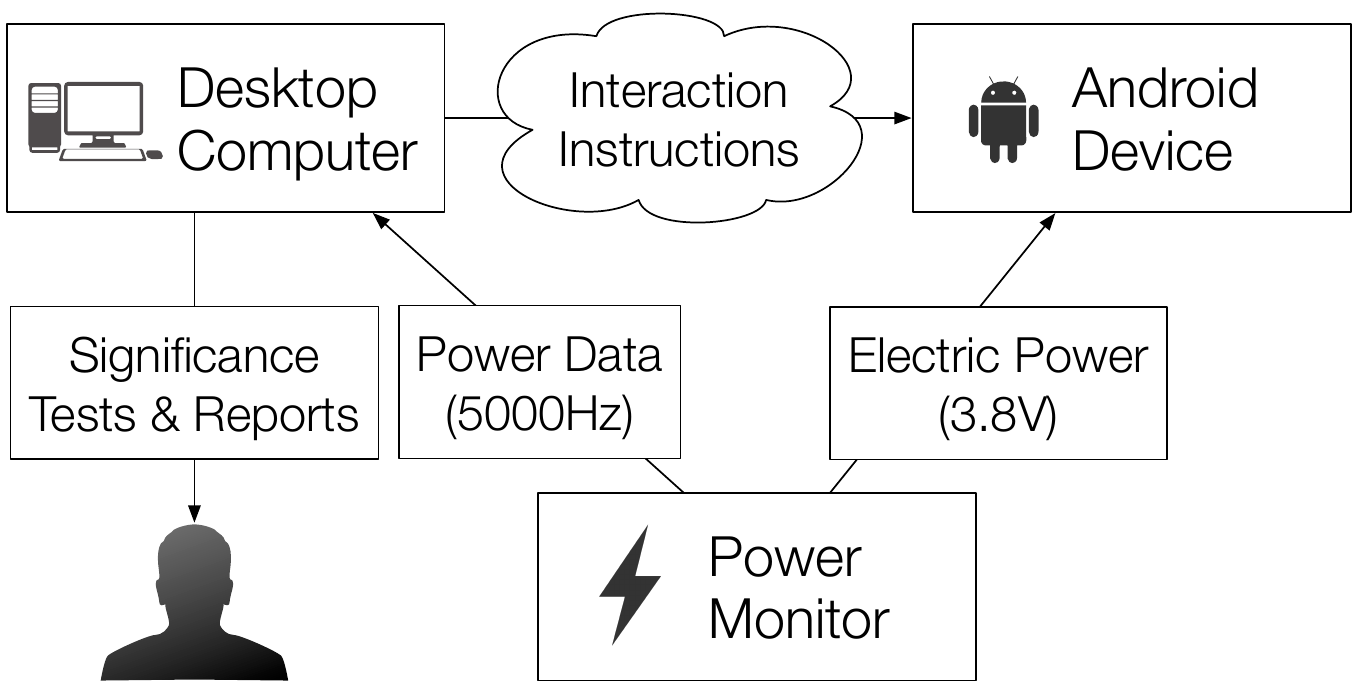}
\caption{Experimentation system to compare UI automation frameworks for
Android.}
\label{fig:arch}
\end{figure}

Our methodology is illustrated in Figure~\ref{fig:arch}. There are three main components:
a desktop computer that serves as controller; a power monitor; and a mobile device
running Android, i.e., the device under test (DUT). The desktop computer sends interaction
instructions to be executed in the mobile device. The power monitor collects
energy consumption data from the mobile device and sends it to the desktop
computer. Finally, the desktop computer analyzes data and generates reports
back to the user.

\subsection{Energy Data Collection} We have adopted a hardware-based approach
to obtain energy measurements. We use Monsoon's \emph{Original Power Monitor}
with the sample rate set to $5000\textup{Hz}$, as used in previous
research~\cite{di2017software,linares2014mining,li2014investigation,li2014making,choudhary2015automated,hindle2015green,banerjee2016automated,banerjee2016debugging}. Measurements are obtained using the \emph{Physalia}
toolset\footnote{Physalia's webpage: \url{https://tqrg.github.io/physalia/} (visited on
\today{}).} -- a Python library to collect energy consumption measurements in
Android devices. It takes care of syncing the beginning and ending of the UI
interaction script with the measurements collected from the power monitor. The
steps described in \emph{Physalia}'s tutorial\footnote{Tutorial's webpage:
\url{https://tqrg.github.io/physalia/monsoon_tutorial} (visited on \today{}).} were followed to remove the
device's battery and connect it directly to the Monsoon's power source using a
constant voltage of $3.8\textup{V}$. This is important to ascertain that we are
collecting reliable energy consumption measurements.

\subsection{Platform}
The choice of the Android platform lies in the fact that it is one of
the most popular operating systems (OS) and is open source. This helps
to understand the underlying system and use a wide range of instrumentation tools.
However, the techniques and ideas discussed in this paper apply to other operating
systems as well.

\subsection{UI Automation Frameworks}
The state-of-the-art UI automation frameworks for Android used in our study are
\emph{Appium}, \emph{UIAutomator}, \emph{PythonUIAutomator},
\emph{AndroidViewClient}, \emph{Espresso}, \emph{Robotium},
\emph{Monkeyrunner}, and \emph{Calabash}. The frameworks were chosen following
a systematic criteria/review: freely available to the community, open source,
featuring a realistic set of interactions, expressed through a human readable
and writable format (e.g., programming language), and used by the mobile
development industry. To assess this last criterion \emph{StackOverflow} and
\emph{Github} were used as proxy. Some frameworks have been discarded
for not complying with this criteria. As an example,
\emph{Ranorex}\footnote{\emph{Ranorex}'s website available at
\url{https://www.ranorex.com} (visited on \today).} is not free to the community
and \emph{RERAN}~\cite{gomez2013reran} is designed to be used with a recording
mechanism. \emph{MonkeyTalk} has not been publicly released after being
acquired by Oracle\footnote{More information about \emph{MonkeyTalk}'s
acquisition: \url{https://www.oracle.com/corporate/acquisitions/cloudmonkey/} (visited on \today{})}, and
\emph{Selendroid} is not ready to be used with the latest Android
SDK\footnote{Running Selendroid
would require changing its source code: \url{https://github.com/selendroid/selendroid/issues/1116}
and \url{https://github.com/selendroid/selendroid/issues/1107} (visited on \today{})}. We decided not to
include UI recording tools since they rely on the underlying frameworks (e.g.,
\emph{Espresso Test Recorder}, \emph{Robotium Recorder}).

Although most frameworks support usage directly through screen
coordinates, we only study the usage by targeting UI components. Usage through
coordinates makes the tests cumbersome to build and maintain, and is not common
practice.

An overview of the features of the frameworks is in
Table~\ref{tab:frameworks_overview}. It also details the frameworks as to
whether the app's \underline{source code} is required, whether it is
\underline{remote script-based}, i.e., simple interaction commands can be sent
in real time to the DUT; \underline{WebView support}, i.e., whether hybrid apps
can also be automated; \underline{compatibility with iOS}, and
\underline{supported programming languages}. The most common languages
supported by these frameworks are \emph{Python} and \emph{Java}.

\begin{table*}
  \caption{Overview of the studied UI automation frameworks}
  \label{tab:frameworks_overview}
 \resizebox{\textwidth}{!}{
\begin{tabular}{lccccccccc}
\hline
Framework                       & Android           & Appium                            & Calabash           & Espresso    & Monkeyrunner  & Python Ui      & Robotium    & UIAutomator   \\
                                & View Client       &                                   &                    &             &               & Automator      &             &               \\
\hline                                                                                                                                                                                                         %
Tap                             & \check{}          & \check{}                          & \check{}           & \check{}    & \check{}      & \check{}       & \check{}    & \check{}       \\
Long Tap                        & \check{}          & \check{}                          & \check{}           & \check{}    & \check{}      & \check{}       & \check{}    & \check{}       \\
Drag And Drop                   & \check{}          & \check{}                          & \check{}           & \no{}       & \check{}      & \check{}       & \check{}    & \check{}       \\
Swipe                           & \check{}          & \check{}                          & \check{}           & \check{}    & \check{}      & \check{}       & \check{}    & \check{}       \\
Pinch \& Spread                 & \no{}             & \no{}                             & \check{}           & \no{}       & \no{}         & \check{}       & \no{}       & \check{}       \\
Back button                     & \check{}          & \check{}                          & \check{}           & \check{}    & \check{}      & \check{}       & \check{}    & \check{}       \\
Input text                      & \check{}          & \check{}                          & \check{}           & \check{}    & \check{}      & \check{}(*)    & \check{}(*) & \check{}(*)    \\
Find by id                      & \check{}          & \check{}                          & \check{}           & \check{}    & \check{}      & \check{}       & \check{}    & \check{}       \\
Find by description             & \check{}          & \check{}                          & \check{}           & \check{}    & \no{}         & \check{}       & \no{}       & \check{}       \\
Find by content                 & \check{}          & \check{}                          & \check{}           & \check{}    & \no{}         & \check{}       & \check{}    & \check{}       \\
\hline                                                                                                                                                                                                                            %
Tested Version                  & 13.2.2            & 1.6.3                             & 0.9.0              & 2.2.2       & \na{}         & 0.3.2          & 5.6.3       & 2.1.2          \\
Min Android SDK                 & All               & Recmd. $\ge$ 17                   & All                & $\ge$ 8     & \na{}         & $\ge$ 18       & $\ge$ 8     & $\ge$ 18       \\
Black Box                       & Yes               & Yes                               & Limited (**)       & No          & Yes           & Yes            & Yes          & Yes              \\
Remote script-based                     & Yes               & Yes                               & Yes                & No          & Yes           & Yes            & No          & No              \\
WebView Support                 & Limited           & Yes                               & Yes                & Yes          & Limited       & Limited        & Yes         & Limited          \\
iOS compatible                  & No                & Yes                               & Yes                & No          & No            & No             & No          & No             \\
BDD support                     & No                & Yes                               & Yes                & Yes         & No            & No             & Yes          & Limited             \\
Integration test                & Yes               & Yes                               & Yes                & No          & No            & Yes            & Yes          & Yes             \\
Language                        & Python            & Any WebDriver                     & Gherkin/Ruby       & Java        & Jython        & Python         & Java        & Java           \\
                                &                   & compatible lang.                  &                    &             &               &                &             &               \\
License                         & Apache 2.0        & Apache 2.0                        & EPL 1.0            & Apache 2.0  & Apache 2.0    & MIT            & Apache 2.0  & Apache 2.0     \\
SOverflow Qns \faStackOverflow  & 164               & 3,147                             & 569                & 292         & 437           & 0              & 1,012       & 438            \\
Github Stars \faGithub          & 540               & 5,514                             & 1,429            & \na\        & \na\          & 719            & 2,165     & \na{}            \\
\hline
\multicolumn{9}{p{0.9\textwidth}}{\scriptsize (*) Although it supports \emph{Input Text}, it does not apply a sequential input of key events. This is more energy efficient but it is more artificial, bypassing real behavior (e.g., auto correct).}\\
\multicolumn{9}{l}{\scriptsize (**) Requires to manually enable Internet permission (\texttt{"android.permission.INTERNET"}).}\\
\hline
\end{tabular}
}
\end{table*}

\subsection{Test cases} For each framework, a script was created for every
interaction that was supported by the framework, totaling 73 scripts. Scripts
were manually and carefully crafted and peer reviewed to ascertain similar
behavior across all frameworks. Essentially, each script calls a specific
method of the framework that mimics the user interaction that we pretend to
study. To minimize overheads from setup tasks (e.g., opening the app, getting
app's UI hierarchy), the method is repeated multiple times: in the case of
\emph{Back Button}, we repeat $200$ times; in the cases of \emph{Swipe},
\emph{Pinch and Spread}, or lookup methods, we repeat $40$ times; in the
remaining interactions, we repeat $10$ times.

\subsection{Setup and Metrics}
We compare the overhead in energy consumption using as baseline the energy
usage of interactions when executed by a human. Baselines for each interaction
were measured by asking two Android users (one female and one male) to execute
the interactions as in the automated scripts.
For instance, in one of the
experiments the participants had to click $200$ times in the \emph{Back Button}.
All interactions were measured except for \emph{Find by id}, \emph{Find
by description}, and \emph{Find by content}, as these are helper methods provided
by the UI automation frameworks and are not applicable to human interactions.

As mentioned above, energy measurements are prone to random variations due to
the nature of the underlying OS. Furthermore, one can also expect errors from
the data collected from a power monitor~\cite{saborido2015impact}. To make sure
energy consumption values are reliable and have enough data to perform
significance tests, each experiment was identically and independently repeated
$30$ times.

Since user interactions often trigger other tasks in a mobile device, tests have to run in a controlled
environment. In other words, we are trying to measure the platform overhead and we don't want the app activity to
interfere with that measurement. Thus, an Android application was developed by the authors for this particular
study. It differs from a real app in the sense that this app is a strategy to prevent any extra work from
being performed by the foreground activity. The main goal is preventing any side-effect from UI interactions,
which in real apps would result in different behaviors, hence compromising measurements. Hence, the app prevents
the propagation of the system's event created by the interaction and no feedback is provided to the user. This
way, experiments only measure the work entailed by frameworks.


The main settings used in the device are listed in Table~\ref{tab:settings}.
Android provides system settings that can be useful to control the system
behavior during experiments. Notifications and alarms were turned off, lock
screen security was disabled, and the ``Don't keep Activities'' setting was
enabled. This last setting destroys every activity as soon as the user leaves
it, erasing the current state of an app\footnote{More about
``Don't Keep Activities'' setting available at:
\url{https://goo.gl/SXkxVy} (visited on \today{}).}.

\begin{table}
  \caption{Android device's system Settings }
  \label{tab:settings}
  \centering
\begin{tabular}{l l}
\hline
Setting                                 & Value\\
\hline
\faSunO\ Adaptive Brightness             & \off\ Manual - $78\%$\\
\faBluetooth\ Bluetooth                  & \off\ Off\\
\faWifi\ WiFi                            & \on\ On\\
Cellular                                 & \off\ No SIM card\\
\faLocationArrow\ Location Services      & \off\ Off\\
\faMobilePhone\ Auto-rotate screen       & \off\ Off - Portrait\\
\faBellO\ Zen mode                       & \on\ On - Total Silence\\
\faLock\ Pin/Pattern Lock Screen         & \off\ Off\\
\faStopCircle\ Don't Keep Activities     & \on\ On\\
\faRefresh\ Account Sync                    & \off\ Off\\
\faAndroid\ Android Version              & 6.0.1\\
\hline
\end{tabular}
\end{table}

WiFi is kept on as a requirement of our experimental setup. The reason lies in
the fact that Android automation frameworks resort to the Android Debug Bridge
(ADB) to communicate with the mobile device. ADB allows to
install/uninstall/open apps, send test data, configure settings, lock/unlock
the device, among other things. By default, it works through USB, which
interferes with energy consumption measurements. Although Android provides
settings to disable USB charging, if the USB port remains connected to the
device, the measurements of energy consumption become obsolete. Fortunately,
ADB can be configured to be used through a WiFi connection, which was leveraged
in this work.

In addition to the energy consumption sources mentioned before, there is another
common one -- the cost of having the device in
idle mode. In this context, we consider idle mode when the device is active
with the settings in Table~\ref{tab:settings} but is not executing any task. In
this mode, the screen is still consuming energy. We calculate the idle cost for
each experiment to assess the effective energy consumption of executing a given
interaction. We measure the idle cost by collecting the energy usage of running
the app for $120$ seconds without any interaction. In addition to the mean
energy consumption, we compare different frameworks using the mean energy
consumption without the corresponding idle cost, calculated as follows:
\begin{equation}
  \bar{x}' = \frac{\sum_{i=1}^{N=30}{(E_i - IdleCost*\Delta t_i)}}{N}
  \label{eq:xbar}
\end{equation}
\noindent where $N$ is the number of times experiments are repeated (30), $E_i$ is the measured energy
consumption for execution $i$, $IdleCost$ is the energy usage per second (i.e., power)
of having the device in idle mode, expressed in watts (W), and $\Delta t_i$ the duration of execution $i$.

After removing idle cost, we compute overhead in a similar fashion as previous work~\cite{greenup}:
\begin{equation}
  Overhead(\%) = (\sfrac{\bar{x}'}{\bar{x}'_{human}}-1)\times 100
  \label{eq:overhead}
\end{equation}
In other words, overhead is the percentage change of the energy consumption of
a framework when compared to the real energy consumption induced by human
interaction.

We also use $\bar{x}'$ to compute the estimated energy consumption for a
single interaction (\emph{Sg}) as follows:
\begin{equation}
  Sg = \sfrac{\bar{{x}}'}{M}
  \label{eq:sg}
\end{equation}
\noindent where $M$
is the number of times the interaction was repeated within the same execution
(e.g., in \emph{Back Button}, $M=200$).


Experiments were executed using an \emph{Apple iMac Mid 2011} with a 2.7GHz
\emph{Intel Core i5} processor, 8GB \emph{DDR3} RAM, and running OS X version
10.11.6. Room temperature was controlled for 24\textdegree C (75\textdegree F).
DUT was a \emph{Nexus 5X} manufactured by \emph{LG}, running Android version
6.0.1. All scripts, mobile app, and data are available in the \emph{Github}
repository of the project\footnote{Project's \emph{Github} repository:
\url{https://github.com/luiscruz/physalia-automators} visited on \today{}.},
which is released with an open source license.
\section{Results}
Next, we report the results obtained in the empirical study.

\subsection{Idle Cost}

\begin{figure}
    \centering
    \includegraphics[width=0.99\linewidth]{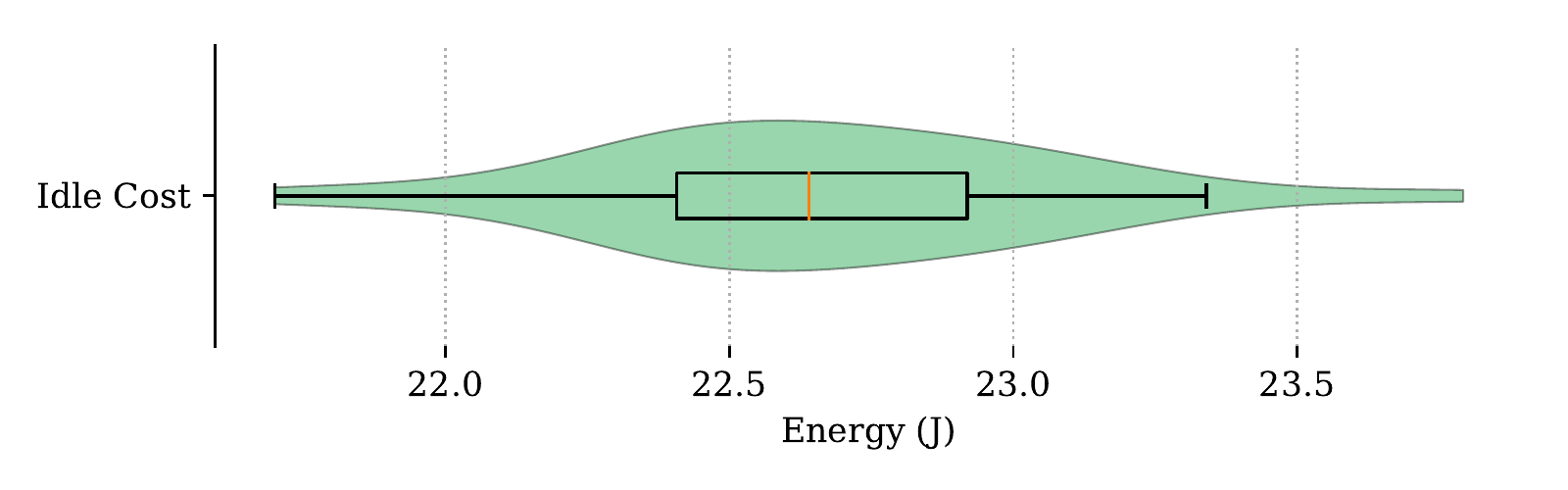}
    
  \caption{Violin plot with distribution of the energy consumption of the app
  during $120$ seconds.}

  \label{fig:idle_cost}
\end{figure}

In a sample of $30$ executions, the mean energy consumption of having the app
open for $120$ seconds without any interaction is $22.67$J. The distribution of
the measurements across the $30$ executions is shown in
Figure~\ref{fig:idle_cost}. This translates into a power consumption of $0.19$W
(in other words, the app consumes $0.19$ joules per second in idle mode). This
value is used in the remaining experiments to factor out idle cost from the
results.

\subsection{\emph{Tap}}
Table~\ref{table_description_tap} presents results for the \emph{Tap}
interaction. Each row in the table describes a framework as a function of the
mean energy consumption ($\bar{x}$), standard deviation of energy consumption
($s$), duration of each execution of the script ($\Delta t$) in seconds, the
mean energy consumption without idle cost ($\bar{{x}}'$, see
Eq.~\ref{eq:xbar}), the estimated energy consumption for a single interaction
($Sg$, see Eq.~\ref{eq:sg}), the Cohen's-$d$ effect size ($d$), the percentage
overhead when compared to the same interaction when executed by a human (as in
Eq.~\ref{eq:overhead}), and the position in the ranking (\#), i.e, the ordinal
position when results are sorted by the average energy consumption, and . With
the exception of the results for \emph{Human} which are placed in the first
row, the table is sorted in alphabetical order for the sake of comparison with
results of other interactions.

\begin{table}
  \caption{Descriptive statistics of \emph{Tap} interaction.}
  \label{table_description_tap}
  \resizebox{1.0\linewidth}{!}{%
  \begin{tabular}{lrrrrrrlr}
\hline
                   &   $\bar{x}$ (J) &   $s$ &   $\Delta t$ (s) &   $\bar{{x}}'$ (J) &   Sg (mJ) &   $d$ & Overhead   &   \# \\
\hline
 \textbf{Human}    &            5.56 &  1.61 &            12.84 &               3.14 &     78.44 &       & ---        &    1 \\
 AndroidViewClient &           19.71 &  0.21 &            42.10 &              11.75 &    293.86 &  8.65 & 274.6\%    &    7 \\
 Appium            &           54.73 &  1.14 &           128.47 &              30.46 &    761.49 & 21.38 & 870.8\%    &    9 \\
 Calabash          &           29.25 &  0.72 &            60.10 &              17.89 &    447.28 & 14.09 & 470.2\%    &    8 \\
 Espresso          &            6.07 &  0.16 &            12.93 &               3.63 &     90.70 &  0.49 & 15.6\%     &    2 \\
 Monkeyrunner      &           18.08 &  1.28 &            49.97 &               8.63 &    215.87 &  4.11 & 175.2\%    &    5 \\
 PythonUiAutomator &            9.15 &  0.54 &            18.93 &               5.57 &    139.32 &  2.24 & 77.6\%     &    4 \\
 Robotium          &           14.59 &  4.00 &            25.63 &               9.74 &    243.57 &  2.11 & 210.5\%    &    6 \\
 UiAutomator       &            7.64 &  0.55 &            17.77 &               4.28 &    107.03 &  1.13 & 36.5\%     &    3 \\
\hline
\end{tabular}

  }
\end{table}

From our experiments, we conclude that \emph{Espresso} is the most energy
efficient framework for \emph{Tap}s, consuming $3.63\textup{J}$ on average
after removing idle cost, while a single \emph{Tap} is estimated to consume
$0.09\textup{J}$. When compared to the human interaction, \emph{Espresso} imposes
an overhead of $16\%$. The least efficient frameworks for a Tap are \emph{Appium},
and \emph{Calabash}, with overheads of $871\%$ and $470\%$, respectively.
Using these frameworks for taps can dramatically affect energy consumption
results.

\begin{figure}
    \centering
    \includegraphics[width=0.99\linewidth]{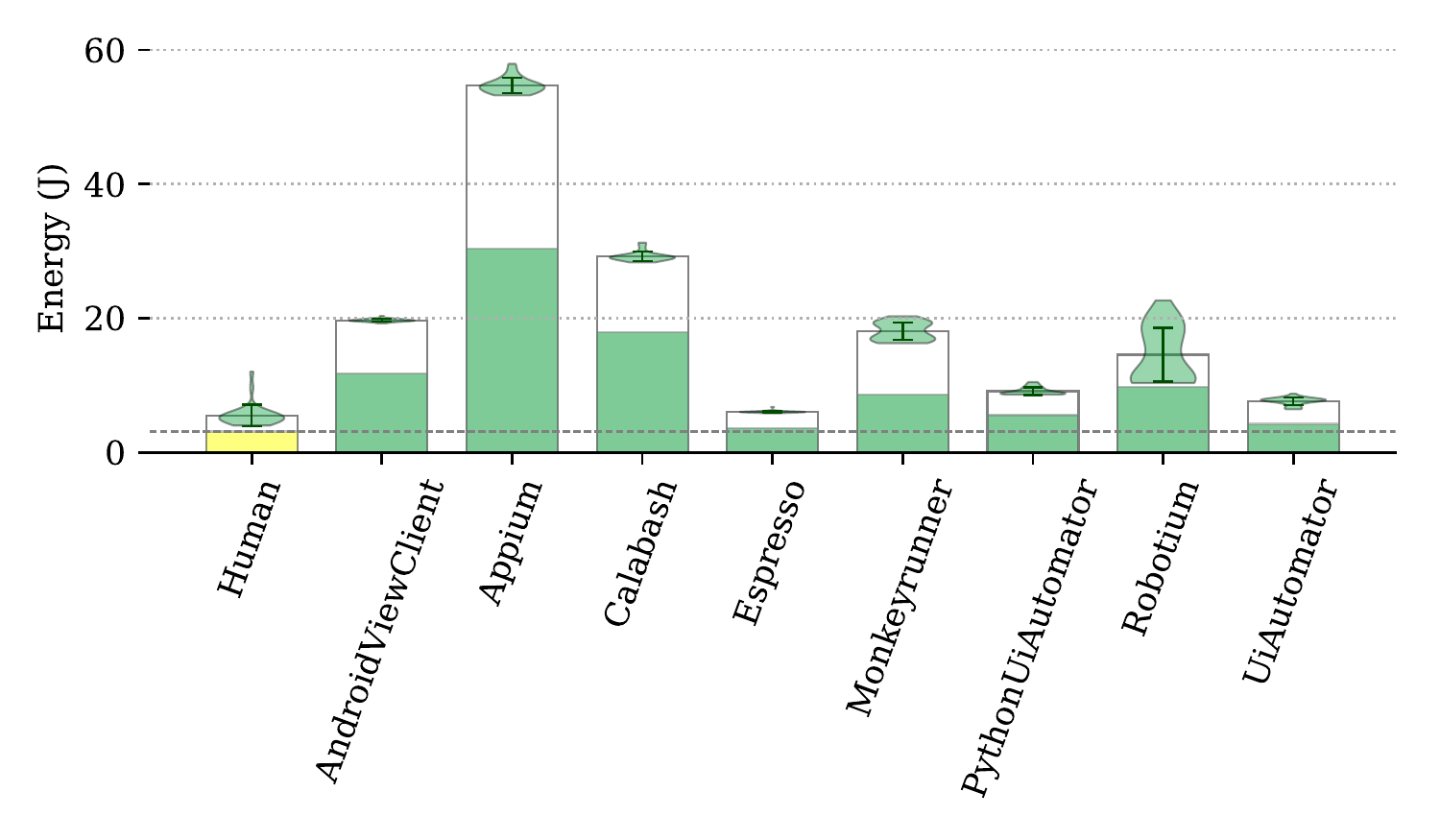}
  \caption{Violin plot of the results for the energy consumption of \emph{Tap}.}
  \label{fig:results_tap}
\end{figure}

A visualization of these results is in Figure~\ref{fig:results_tap}. The height
of each white bar shows the mean energy consumption for the framework. The
height of each green or yellow bar represents the energy consumption without
the idle cost. The yellow bar and the dashed horizontal line highlight the
baseline energy consumption. In addition, it shows a violin plot with the
probability density of data using rotated kernel density plots. The violin
plots provide a visualization of the distribution, allowing to compare results
regarding shape, location, and scale. This is useful to assess whether data can
be modeled with a normal distribution, and compare the standard deviations of
the measurements in different frameworks.

\subsection{\emph{Long Tap}}

\begin{figure}
    \centering
    \includegraphics[width=0.99\linewidth]{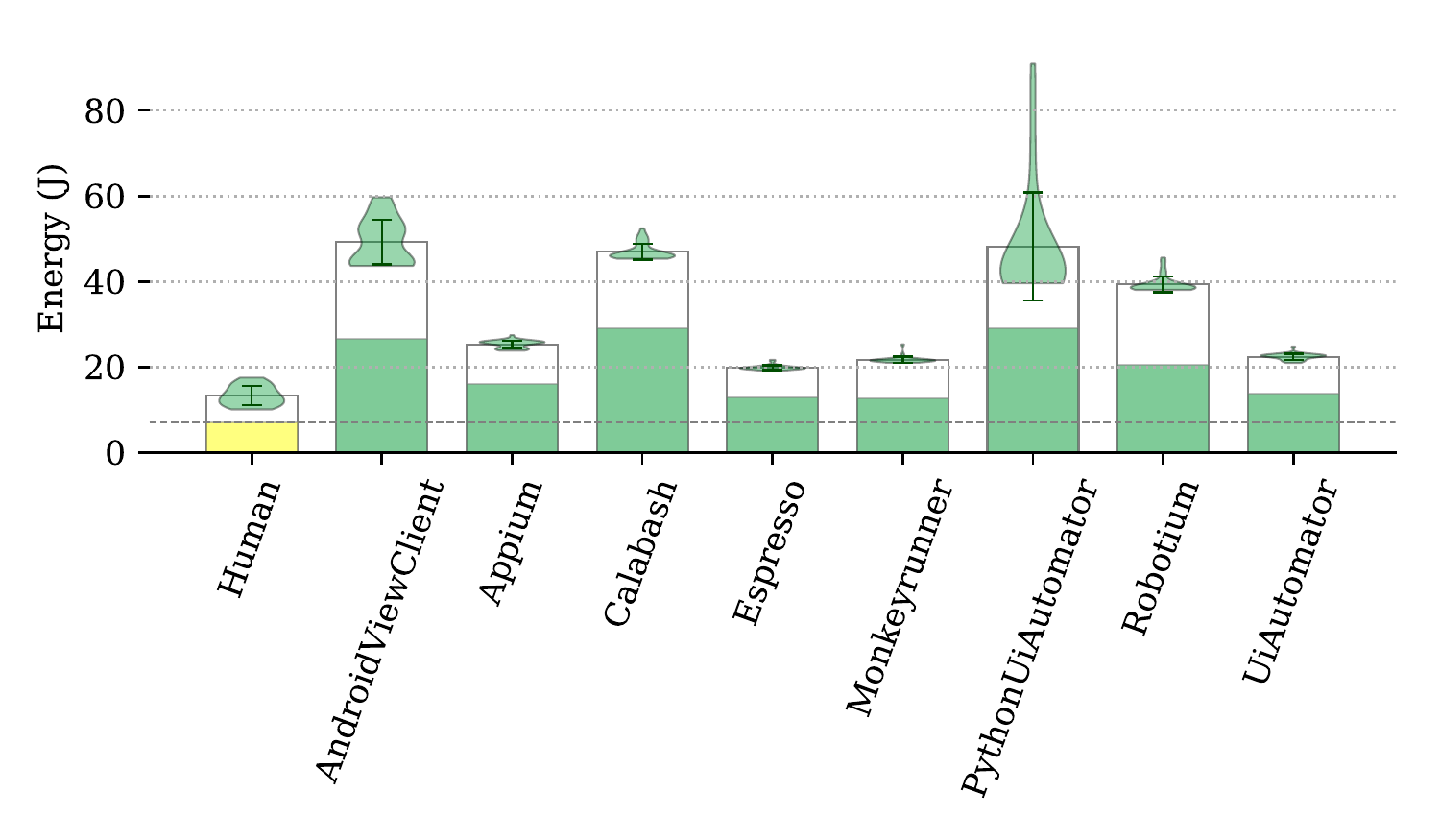}
  \caption{Violin plot of the results for energy consumption of \emph{Long Tap}.}
  \label{fig:results_long_tap}
\end{figure}

Results for the interaction \emph{Long Tap} are in
Table~\ref{table_description_long_tap} and Figure~\ref{fig:results_long_tap}.
\emph{Monkeyrunner} and \emph{Espresso} are the most efficient frameworks,
with overheads of $77\%$ ($\bar{{x}}' = 12.60\textup{J}$) and $81\%$
($\bar{{x}}' = 12.88\textup{J}$), respectively. \emph{PythonUIAutomator} and
\emph{Calabash} are the most inefficient (overhead over $300\%$).

A remarkable observation is the efficiency of
\emph{Appium}'s \emph{Long Tap} ($Sg=0.40\textup{J}$) when compared to its
regular \emph{Tap} ($Sg=0.76\textup{J}$). Common sense would let us expect
\emph{Tap} to spend less energy than \emph{Long Tap}, but that is not the case.
This happens because \emph{Appium}'s usage of \emph{Long Tap} requires a manual
instantiation of a \texttt{TouchAction} object, while \emph{Tap} creates it
internally. Although creating such object makes code less readable, the
advantage is that it can be reused for the following interactions, making a
more efficient use of resources.

\begin{table}
  \caption{Descriptive statistics of \emph{Long Tap} interaction.}
  \label{table_description_long_tap}
  \resizebox{1.0\linewidth}{!}{%
  \begin{tabular}{lrrrrrrlr}
\hline
                   &   $\bar{x}$ (J) &   $s$ &   $\Delta t$ (s) &   $\bar{{x}}'$ (J) &   Sg (mJ) &   $d$ & Overhead   &   \# \\
\hline
 \textbf{Human}    &           13.33 &  2.21 &            32.86 &               7.12 &    177.92 &       & ---        &    1 \\
 AndroidViewClient &           49.18 &  5.24 &           119.21 &              26.66 &    666.40 &  4.45 & 274.6\%    &    7 \\
 Appium            &           25.34 &  0.86 &            49.60 &              15.96 &    399.08 &  7.75 & 124.3\%    &    5 \\
 Calabash          &           46.96 &  1.81 &            94.27 &              29.14 &    728.57 & 12.44 & 309.5\%    &    9 \\
 Espresso          &           19.87 &  0.54 &            37.00 &              12.88 &    321.94 &  6.20 & 80.9\%     &    3 \\
 Monkeyrunner      &           21.68 &  0.74 &            48.07 &              12.60 &    315.04 &  5.57 & 77.1\%     &    2 \\
 PythonUiAutomator &           48.19 & 12.63 &           101.13 &              29.08 &    727.02 &  2.70 & 308.6\%    &    8 \\
 Robotium          &           39.35 &  1.82 &            99.97 &              20.46 &    511.40 &  8.71 & 187.4\%    &    6 \\
 UiAutomator       &           22.39 &  0.75 &            45.40 &              13.81 &    345.20 &  6.90 & 94.0\%     &    4 \\
\hline
\end{tabular}

  }
\end{table}

\subsection{\emph{Drag and Drop}}
\label{sec:dragndrop}

\begin{figure}
    \centering
    \includegraphics[width=0.99\linewidth]{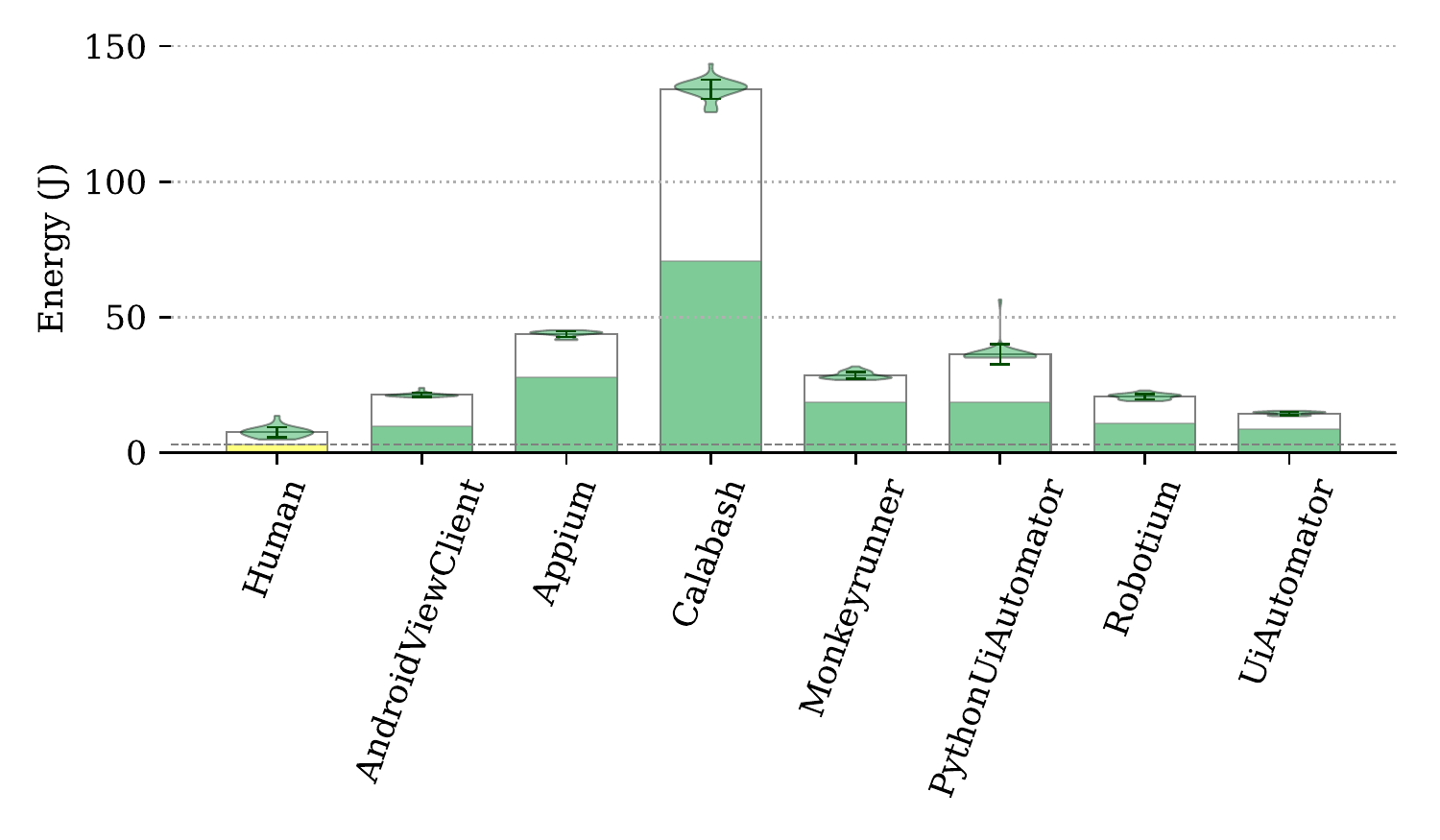}
  \caption{Violin plot of the results for energy consumption of \emph{Drag and Drop}.}
  \label{fig:results_dragndrop}

\end{figure}

Results for the interaction \emph{Drag and Drop} are in
Table~\ref{table_description_dragndrop} and
Figure~\ref{fig:results_dragndrop}. \emph{UIAutomator} is the best testing
framework with an overhead of $185\%$ ($\bar{{x}}' = 14.48\textup{J}$).
\emph{Espresso} is not included in the experiments since \emph{Drag and Drops}
are not supported. The most energy greedy framework is \emph{Calabash} with an
overhead of $2193\%$. When compared to \emph{UIAutomator}, one \emph{Drag and
Drop} with \emph{Calabash} is equivalent to more than 11 \emph{Drag and Drops}.
Hence, \emph{Calabash} should be avoided for energy measurements that include
\emph{Drag and Drops}.

\begin{table}
  \caption{Descriptive statistics of \emph{Drag and Drop} interaction.}
  \label{table_description_dragndrop}
  \resizebox{1.0\linewidth}{!}{%
  \begin{tabular}{lrrrrrrlr}
\hline
                   &   $\bar{x}$ (J) &   $s$ &   $\Delta t$ (s) &   $\bar{{x}}'$ (J) &   Sg (mJ) &   $d$ & Overhead   &   \# \\
\hline
 \textbf{Human}    &            7.55 &  1.91 &            23.70 &               3.08 &     76.90 &       & ---        &    1 \\
 AndroidViewClient &           21.31 &  0.76 &            62.15 &               9.57 &    239.24 &  6.67 & 211.1\%    &    3 \\
 Appium            &           43.71 &  1.14 &            85.00 &              27.65 &    691.27 & 19.69 & 798.9\%    &    7 \\
 Calabash          &          134.08 &  3.55 &           336.33 &              70.53 &   1763.27 & 26.79 & 2193.0\%   &    8 \\
 Monkeyrunner      &           28.50 &  1.29 &            52.97 &              18.49 &    462.22 & 17.96 & 501.1\%    &    5 \\
 PythonUiAutomator &           36.30 &  3.77 &            93.53 &              18.62 &    465.56 &  5.54 & 505.4\%    &    6 \\
 Robotium          &           20.63 &  1.02 &            52.17 &              10.77 &    269.29 &  7.75 & 250.2\%    &    4 \\
 UiAutomator       &           14.48 &  0.63 &            30.27 &               8.76 &    219.02 &  8.19 & 184.8\%    &    2 \\
\hline
\end{tabular}
}
\end{table}

\subsection{\emph{Swipe}}

\begin{figure}
    \centering
    \includegraphics[width=0.99\linewidth]{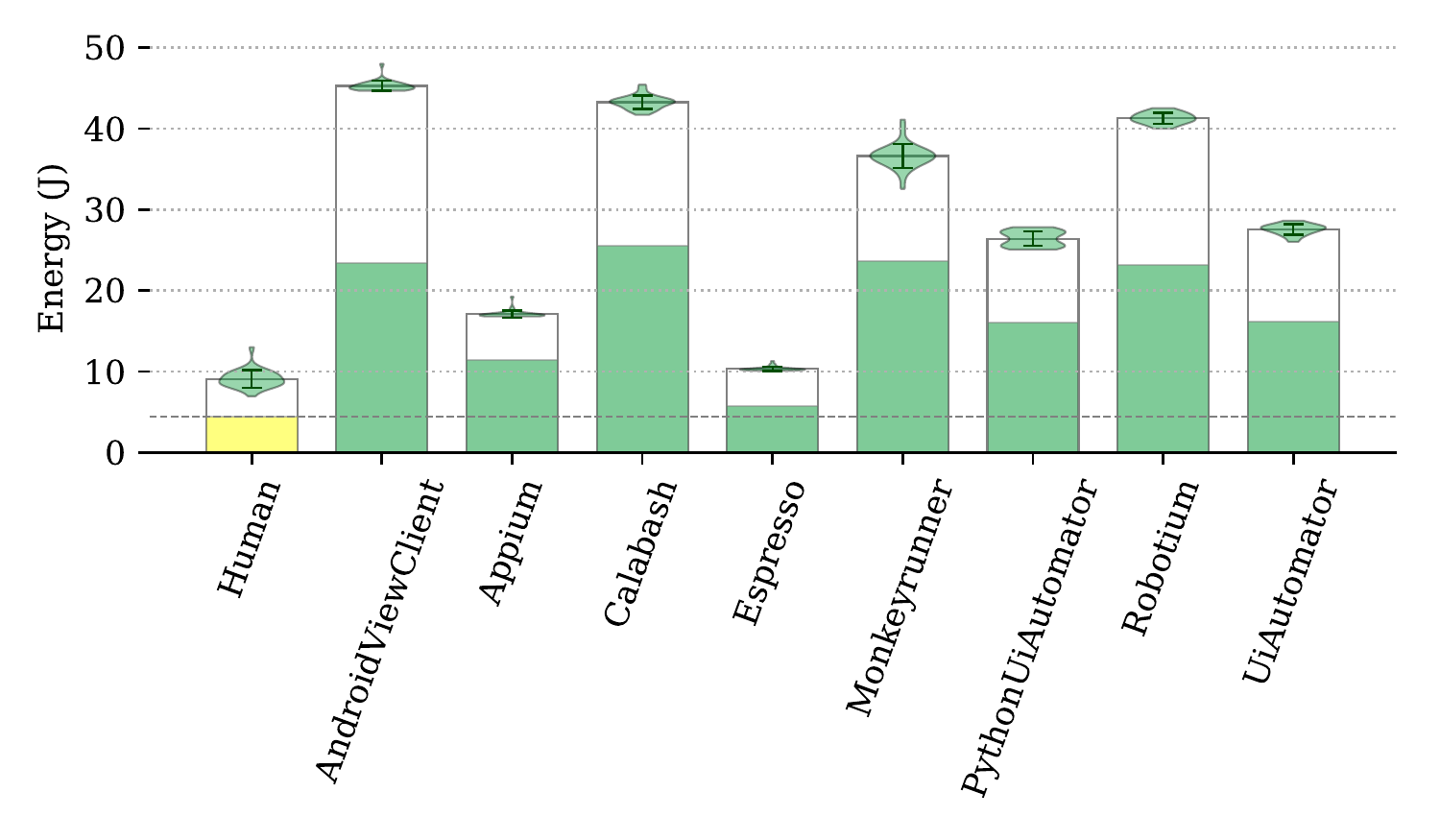}
  \caption{Violin plot of the results for energy consumption of \emph{Swipe}.}
  \label{fig:results_swipe}

\end{figure}

Results for the interaction \emph{Swipe} are presented in
Table~\ref{table_description_swipe} and Figure~\ref{fig:results_swipe}.
\emph{Espresso} is the best framework with an overhead of $29\%$, while
\emph{Robotium}, \emph{AndroidViewClient}, \emph{Monkeyrunner}, and \emph{Calabash}
are the most energy greedy with similar overheads, above $400\%$.

\begin{table}
  \caption{Descriptive statistics of \emph{Swipe} interaction.}
  \label{table_description_swipe}
  \resizebox{1.0\linewidth}{!}{%
  \begin{tabular}{lrrrrrrlr}
\hline
                   &   $\bar{x}$ (J) &   $s$ &   $\Delta t$ (s) &   $\bar{{x}}'$ (J) &   Sg (mJ) &   $d$ & Overhead   &   \# \\
\hline
 \textbf{Human}    &            9.11 &  1.09 &            24.48 &               4.48 &     56.05 &       & ---        &    1 \\
 AndroidViewClient &           45.29 &  0.62 &           115.87 &              23.39 &    292.41 & 27.93 & 421.7\%    &    7 \\
 Appium            &           17.09 &  0.46 &            30.00 &              11.42 &    142.80 & 11.53 & 154.8\%    &    3 \\
 Calabash          &           43.27 &  0.81 &            93.73 &              25.56 &    319.46 & 27.21 & 469.9\%    &    9 \\
 Espresso          &           10.35 &  0.26 &            24.10 &               5.79 &     72.43 &  2.51 & 29.2\%     &    2 \\
 Monkeyrunner      &           36.63 &  1.49 &            68.67 &              23.65 &    295.69 & 25.16 & 427.5\%    &    8 \\
 PythonUiAutomator &           26.42 &  0.91 &            54.60 &              16.11 &    201.32 & 14.05 & 259.1\%    &    4 \\
 Robotium          &           41.30 &  0.67 &            96.00 &              23.16 &    289.46 & 26.85 & 416.4\%    &    6 \\
 UiAutomator       &           27.56 &  0.65 &            60.13 &              16.20 &    202.49 & 19.93 & 261.2\%    &    5 \\
\hline
\end{tabular}
}
\end{table}

\subsection{\emph{Pinch and Spread}}

\begin{figure}
    \centering
    \includegraphics[width=0.99\linewidth]{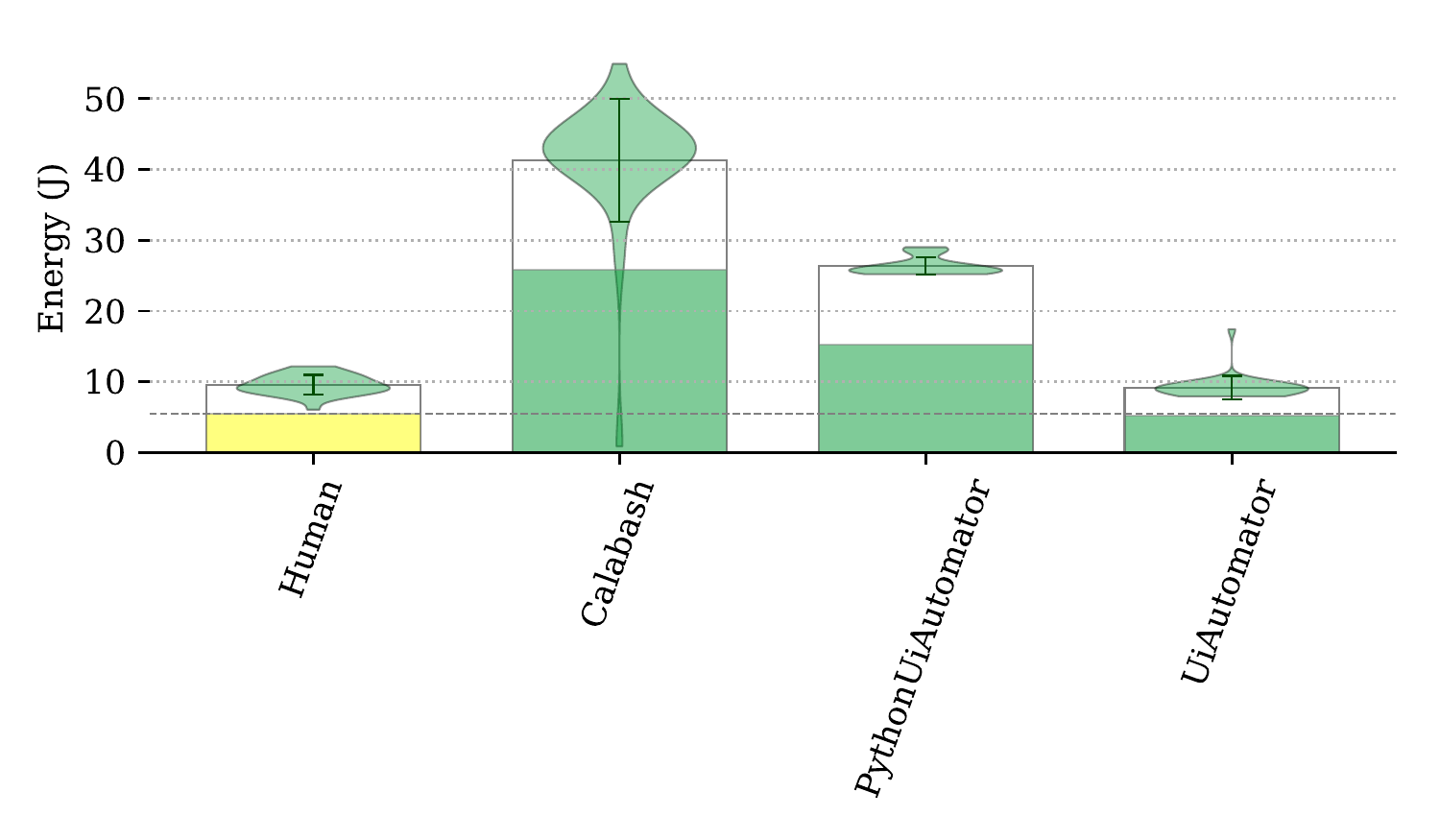}
  \caption{Violin plot of the results for energy consumption of \emph{Pinch and Spread}.}
  \label{fig:results_pinch_and_spread}

\end{figure}

Results for the interaction \emph{Pinch and Spread} are presented in
Table~\ref{table_description_pinch_and_spread} and
Figure~\ref{fig:results_pinch_and_spread}. Although this interaction is widely
used in mobile applications for features such as zoom in and out, only
\emph{Calabash}, \emph{PythonUIAutomator}, and \emph{UIAutomator} support it
out of the box. \emph{UIAutomator} is the most efficient framework, spending
less energy than the equivalent interaction performed by a human ($-5\%$). The
remaining frameworks, \emph{PythonUiAutomator} and \emph{Calabash} were not as
efficient, providing overheads of $181\%$ and $374\%$, respectively.

\begin{table}
  \caption{Descriptive statistics of \emph{Pinch and Spread} interaction.}
  \label{table_description_pinch_and_spread}
  \resizebox{1.0\linewidth}{!}{%
  \begin{tabular}{lrrrrrrlr}
\hline
                      &   $\bar{x}$ (J) &   $s$ &   $\Delta t$ (s) &   $\bar{{x}}'$ (J) &   Sg (mJ) &   $d$ & Overhead   &   \# \\
\hline
 Human                &            9.59 &  1.37 &            21.91 &               5.45 &     68.10 &       & ---        &    2 \\
 Calabash             &           41.31 &  8.66 &            81.93 &              25.83 &    322.83 &  4.82 & 374.0\%    &    4 \\
 PythonUiAutomator    &           26.39 &  1.23 &            58.77 &              15.29 &    191.09 &  9.59 & 180.6\%    &    3 \\
 \textbf{UiAutomator} &            9.19 &  1.66 &            21.23 &               5.17 &     64.67 & -0.21 & -5.0\%     &    1 \\
\hline
\end{tabular}
}
\end{table}

\subsection{\emph{Back Button}}

\begin{figure}
    \centering
    \includegraphics[width=0.99\linewidth]{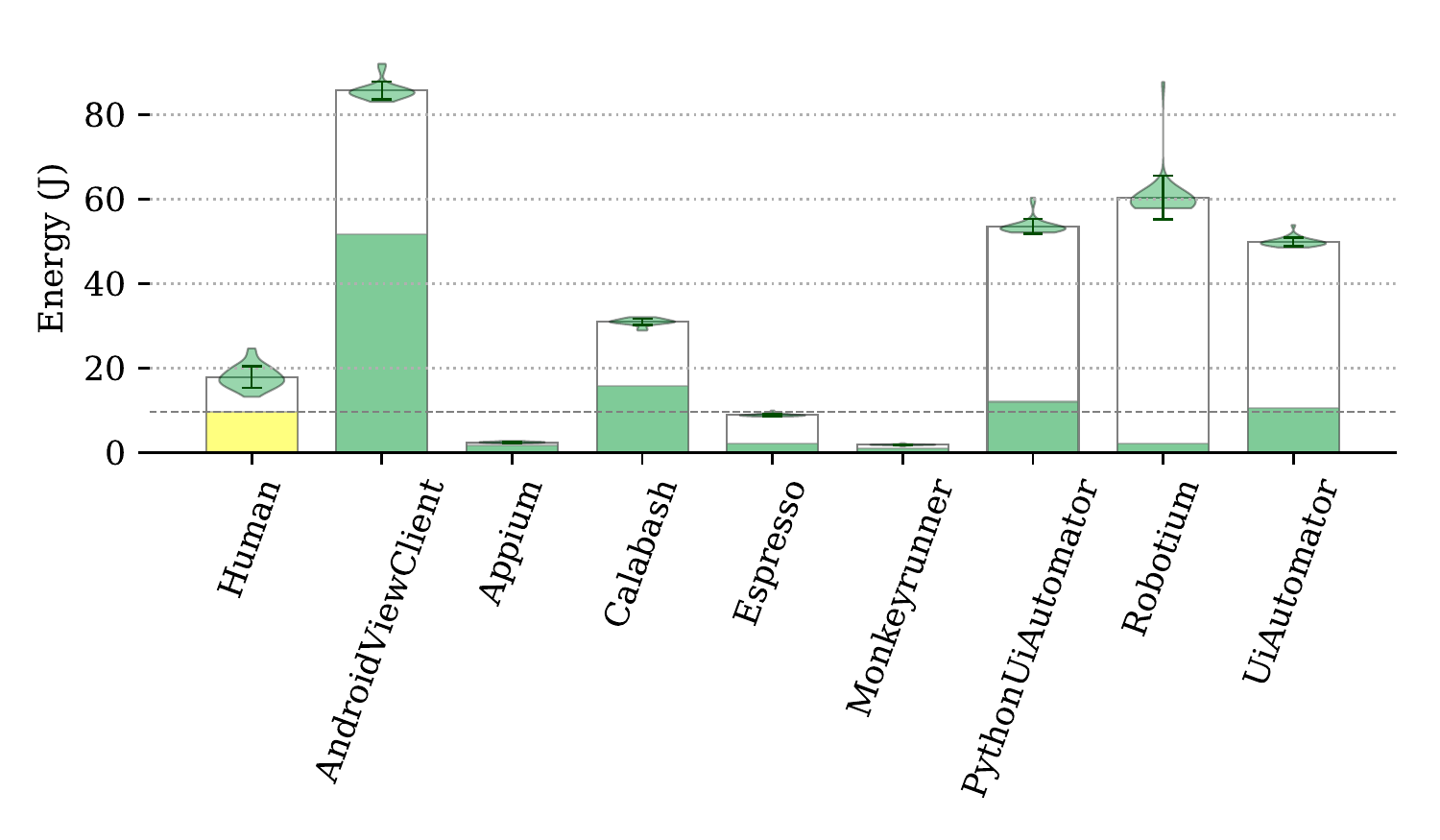}
  \caption{Violin plot of the results for energy consumption of \emph{Back Button}.}
  \label{fig:results_back_button}

\end{figure}

Results for the interaction \emph{Back Button} are presented in
Table~\ref{table_description_back_button} and
Figure~\ref{fig:results_back_button}. In this case, human interaction was
considerably less efficient than most frameworks, being ranked fifth on the
list. The main reason for this is that frameworks do not realistically mimic the
\emph{Back Button} interaction. When the user presses the back button, the system
produces an input event and a vibration or haptic feedback simultaneously.
However, frameworks simply produce the event. Thus, results are not comparable
with the human interaction. Still, \emph{AndroidViewClient} provided an
overhead of $440\%$, being the most inefficient framework.

Another remarkable result was that \emph{Robotium}, despite being energy
efficient after removing idle cost, it is the slowest framework. Thus, it is
likely that \emph{Robotium} is using a conservative approach to generate events
in the device: it suspends the execution to wait for the back button event to
take effect in the app.

\begin{table}
  \caption{Descriptive statistics of \emph{Back Button} interaction.}
  \label{table_description_back_button}
  \resizebox{1.0\linewidth}{!}{%
  \begin{tabular}{lrrrrrrlr}
\hline
                       &   $\bar{x}$ (J) &   $s$ &   $\Delta t$ (s) &   $\bar{{x}}'$ (J) &   Sg (mJ) &   $d$ & Overhead   &   \# \\
\hline
 Human                 &           17.90 &  2.56 &            43.94 &               9.60 &     47.98 &       & ---        &    5 \\
 AndroidViewClient     &           85.75 &  2.11 &           179.73 &              51.79 &    258.94 & 17.79 & 439.7\%    &    9 \\
 Appium                &            2.43 &  0.17 &             3.33 &               1.80 &      9.01 & -8.33 & -81.2\%    &    2 \\
 Calabash              &           30.95 &  0.77 &            80.57 &              15.73 &     78.63 &  5.99 & 63.9\%     &    8 \\
 Espresso              &            8.89 &  0.29 &            35.17 &               2.25 &     11.25 & -7.66 & -76.6\%    &    4 \\
 \textbf{Monkeyrunner} &            1.84 &  0.12 &             4.07 &               1.08 &      5.38 & -9.13 & -88.8\%    &    1 \\
 PythonUiAutomator     &           53.62 &  1.78 &           220.03 &              12.04 &     60.20 &  1.68 & 25.5\%     &    7 \\
 Robotium              &           60.44 &  5.18 &           308.10 &               2.22 &     11.10 & -1.88 & -76.9\%    &    3 \\
 UiAutomator           &           49.87 &  1.03 &           208.20 &              10.53 &     52.64 &  0.83 & 9.7\%      &    6 \\
\hline
\end{tabular}
}
\end{table}

\subsection{\emph{Input Text}}
\label{sec:input_text}

\begin{figure}
    \centering
    \includegraphics[width=0.99\linewidth]{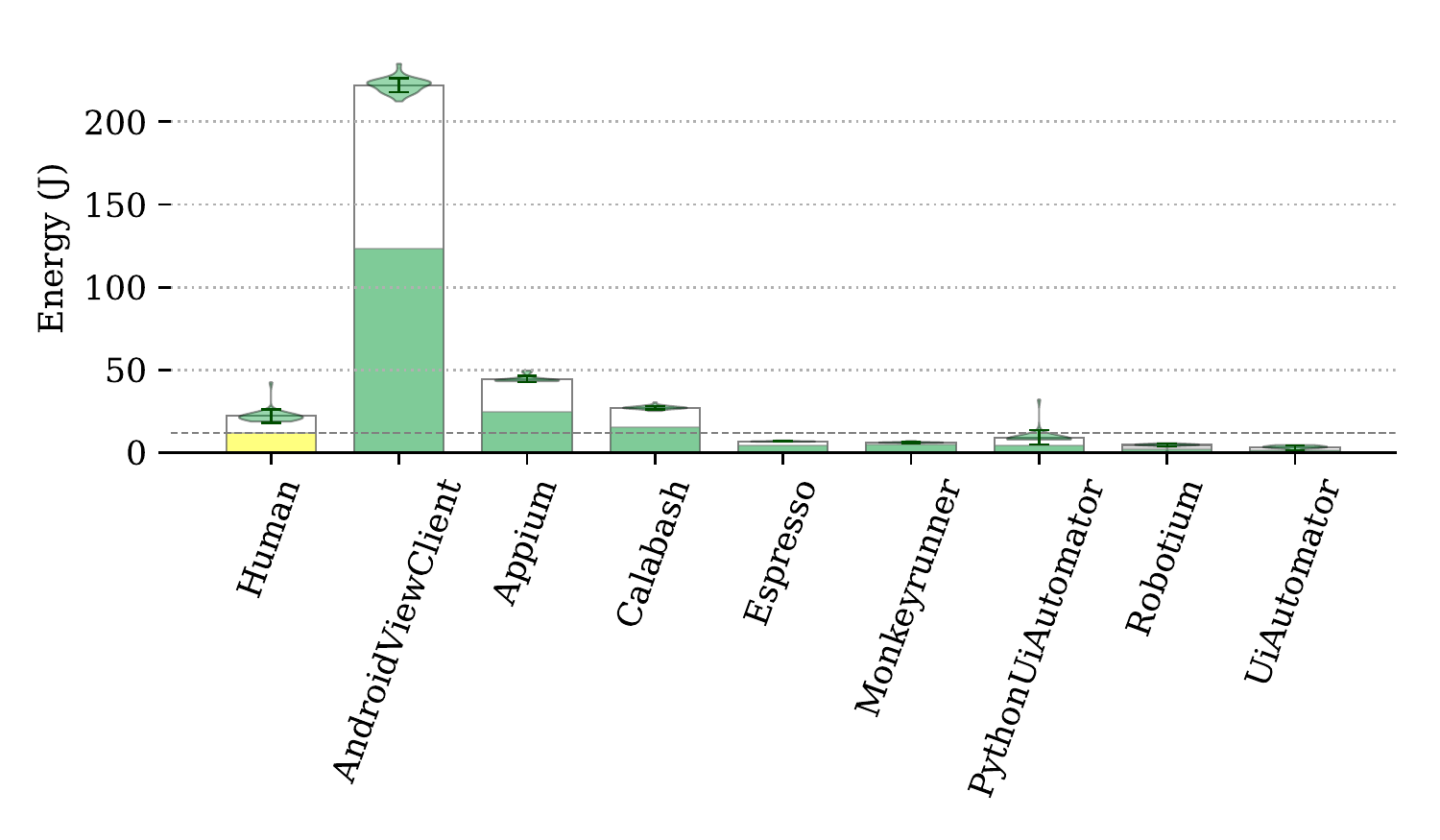}
  \caption{Violin plot of the results for energy consumption of \emph{Input Text}.}
  \label{fig:results_input_text}

\end{figure}

Results for the interaction \emph{Input Text} are presented in
Table~\ref{table_description_input_text} and
Figure~\ref{fig:results_input_text}. Each iteration of \emph{Input Text}
consists in writing a $17$-character sentence in a text field and then clearing
it all back to the initial state. Thus, the value for a single interaction
(\emph{Sg}) is the energy spent when this sequence of events is executed, but
can hardly be extrapolated for other input interactions.

\emph{UIAutomator} is the framework with the lowest energy consumption
($\bar{x}' = 1.42\textup{J}$). The human interaction spends more energy than
most frameworks. The reason behind this is that frameworks have a different way
to deal with text input. Most frameworks generate a sequence of events that
will generate the given sequence of characters. On the contrary, the human
interaction resorts to the system keyboard to generate this sequence. Thus the
system has to process a sequence of taps and match it to the right character
event. There are even other frameworks, namely \emph{UIAutomator},
\emph{PythonUIAutomator}, and \emph{Robotium}, that, as showed in the overview
of Table~\ref{tab:frameworks_overview}, implement \emph{Input Text} more
artificially. Instead of generating the sequence of events, they directly
change the content of the text field. This is more efficient but bypasses system and
application behavior -- e.g., automatic text correction features.

Results showed that the \emph{AndroidViewClient} is very inefficient and its
overhead ($936\%$) is not negligible when measuring the energy consumption of
mobile apps.

\begin{table}
  \caption{Descriptive statistics of \emph{Input Text} interaction.}
  \label{table_description_input_text}
  \resizebox{1.0\linewidth}{!}{%
  \begin{tabular}{lrrrrrrlr}
\hline
                      &   $\bar{x}$ (J) &   $s$ &   $\Delta t$ (s) &   $\bar{{x}}'$ (J) &   Sg (mJ) &   $d$ & Overhead   &   \# \\
\hline
 Human                &           22.11 &  4.06 &            54.09 &              11.89 &   1189.37 &       & ---        &    6 \\
 AndroidViewClient    &          222.08 &  4.31 &           523.37 &             123.18 &  12318.21 & 27.68 & 935.7\%    &    9 \\
 Appium               &           44.43 &  1.89 &           105.27 &              24.54 &   2453.84 &  4.62 & 106.3\%    &    8 \\
 Calabash             &           27.14 &  1.03 &            62.40 &              15.35 &   1534.70 &  1.40 & 29.0\%     &    7 \\
 Espresso             &            6.83 &  0.18 &            14.03 &               4.18 &    417.96 & -3.45 & -64.9\%    &    3 \\
 Monkeyrunner         &            6.18 &  0.29 &             8.03 &               4.67 &    466.58 & -3.21 & -60.8\%    &    5 \\
 PythonUiAutomator    &            9.16 &  4.35 &            25.37 &               4.37 &    436.83 & -2.07 & -63.3\%    &    4 \\
 Robotium             &            4.64 &  0.86 &            12.50 &               2.27 &    227.34 & -4.26 & -80.9\%    &    2 \\
 \textbf{UiAutomator} &            2.93 &  1.39 &             8.00 &               1.42 &    142.02 & -4.62 & -88.1\%    &    1 \\
\hline
\end{tabular}
}
\end{table}

\subsection{\emph{Find by id}}

\begin{figure}
    \centering
    \includegraphics[width=0.99\linewidth]{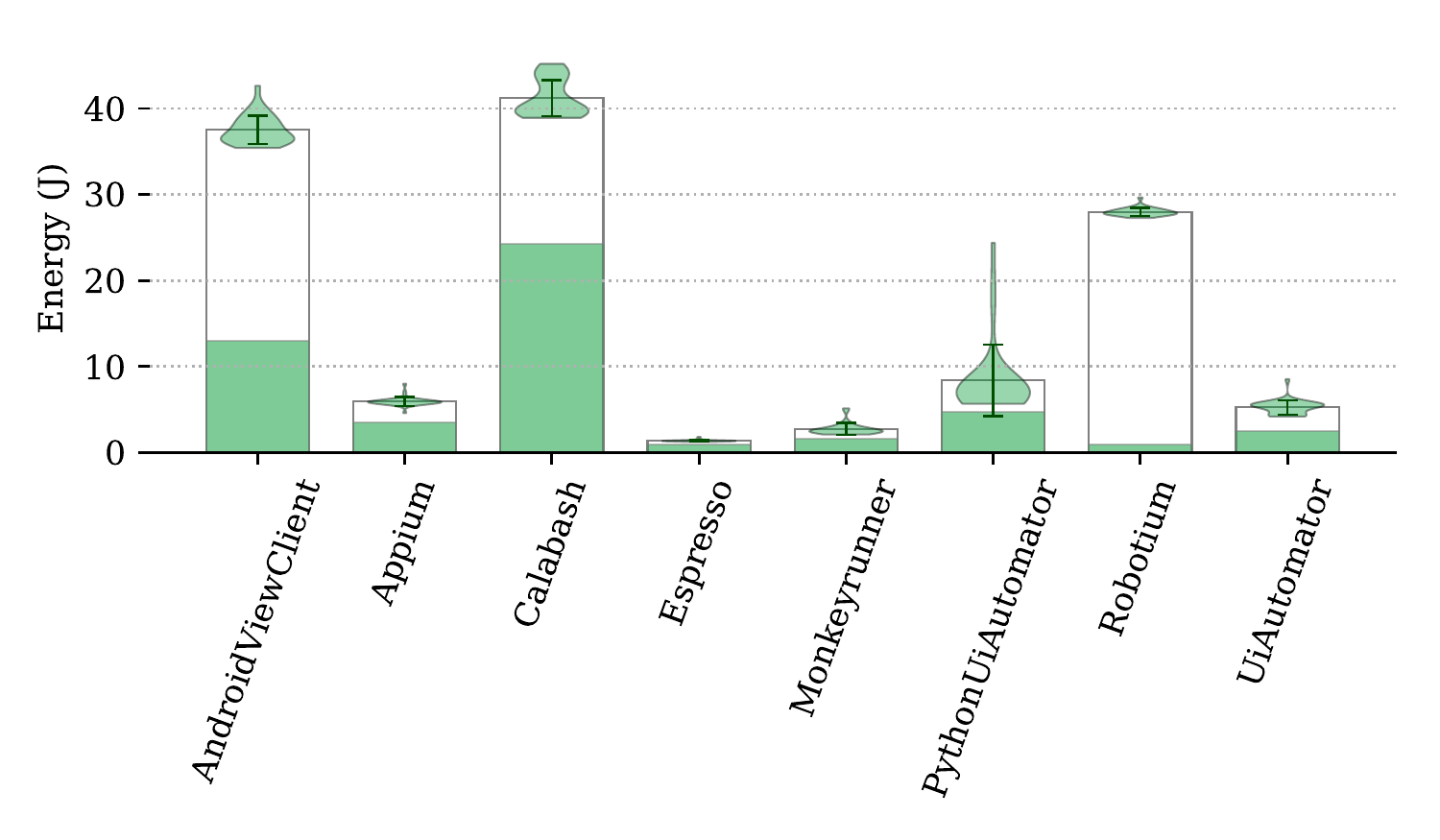}
  \caption{Violin plot of the results for energy consumption of \emph{Find by id}.}
  \label{fig:results_find_by_id}

\end{figure}

Results for the task \emph{Find by id} are presented in
Table~\ref{table_description_find_by_id} and
Figure~\ref{fig:results_find_by_id}. \emph{Find by id} is a method that looks
up for a UI component that has the given \emph{id}. It does not mimic any user
interaction but it is necessary to create interaction scripts. Methods
\emph{Find by description} and \emph{Find by content} are used to achieve the
same objective. For this reason, we do not report the consumption of a human
interaction in these cases.

For the sake of consistency with previous cases, we report tables and
figures in the same fashion. However, we consider that the overall cost of
energy consumption (without removing idle cost) should not be discarded.

\emph{Robotium} is the most energy efficient, with an energy consumption
without idle cost of $0.94J$. However, if we consider idle cost,
\emph{Robotium} is amongst the most energy greedy frameworks (after
\emph{Calabash} and \emph{AndroidViewClient}). It has an overall energy
consumption of $27.97J$. When considering idle cost, \emph{Espresso} is the
most energy efficient framework.

This difference lies in the mechanism adopted by frameworks to deal with UI
changes. After user interaction, the UI is expected to change and the status of
the UI can become obsolete. Thus, frameworks need to wait until the changes the
UI are complete. Results show that \emph{Robotium} uses a mechanism based on
suspending the execution to make sure the UI is up to date. On the other hand,
Espresso uses a different heuristic, which despite spending more energy on
computation tasks, it does not require the device to spend energy while waiting.

\begin{table}
  \caption{Descriptive statistics of \emph{Find by id} interaction.}
  \label{table_description_find_by_id}
  \resizebox{1.0\linewidth}{!}{%
  \begin{tabular}{lrrrrrr}
\hline
                   &   $\bar{x}$ (J) &   $s$ &   $\Delta t$ (s) &   $\bar{{x}}'$ (J) &   Sg (mJ) &   Rank \\
\hline
 AndroidViewClient &           37.52 &  1.64 &           129.91 &              12.97 &     46.34 &      7 \\
 Appium            &            5.94 &  0.51 &            12.73 &               3.53 &     12.62 &      5 \\
 Calabash          &           41.20 &  2.08 &            89.63 &              24.26 &     86.65 &      8 \\
 Espresso          &            1.37 &  0.11 &             2.03 &               0.99 &      3.54 &      2 \\
 Monkeyrunner      &            2.74 &  0.70 &             6.13 &               1.58 &      5.66 &      3 \\
 PythonUiAutomator &            8.42 &  4.16 &            19.63 &               4.71 &     16.81 &      6 \\
 \textbf{Robotium} &           27.97 &  0.46 &           143.03 &               0.94 &      3.37 &      1 \\
 UiAutomator       &            5.26 &  0.84 &            14.33 &               2.55 &      9.11 &      4 \\
\hline
\end{tabular}
}
\end{table}

\subsection{\emph{Find by description}}

\begin{figure}
    \centering
    \includegraphics[width=0.99\linewidth]{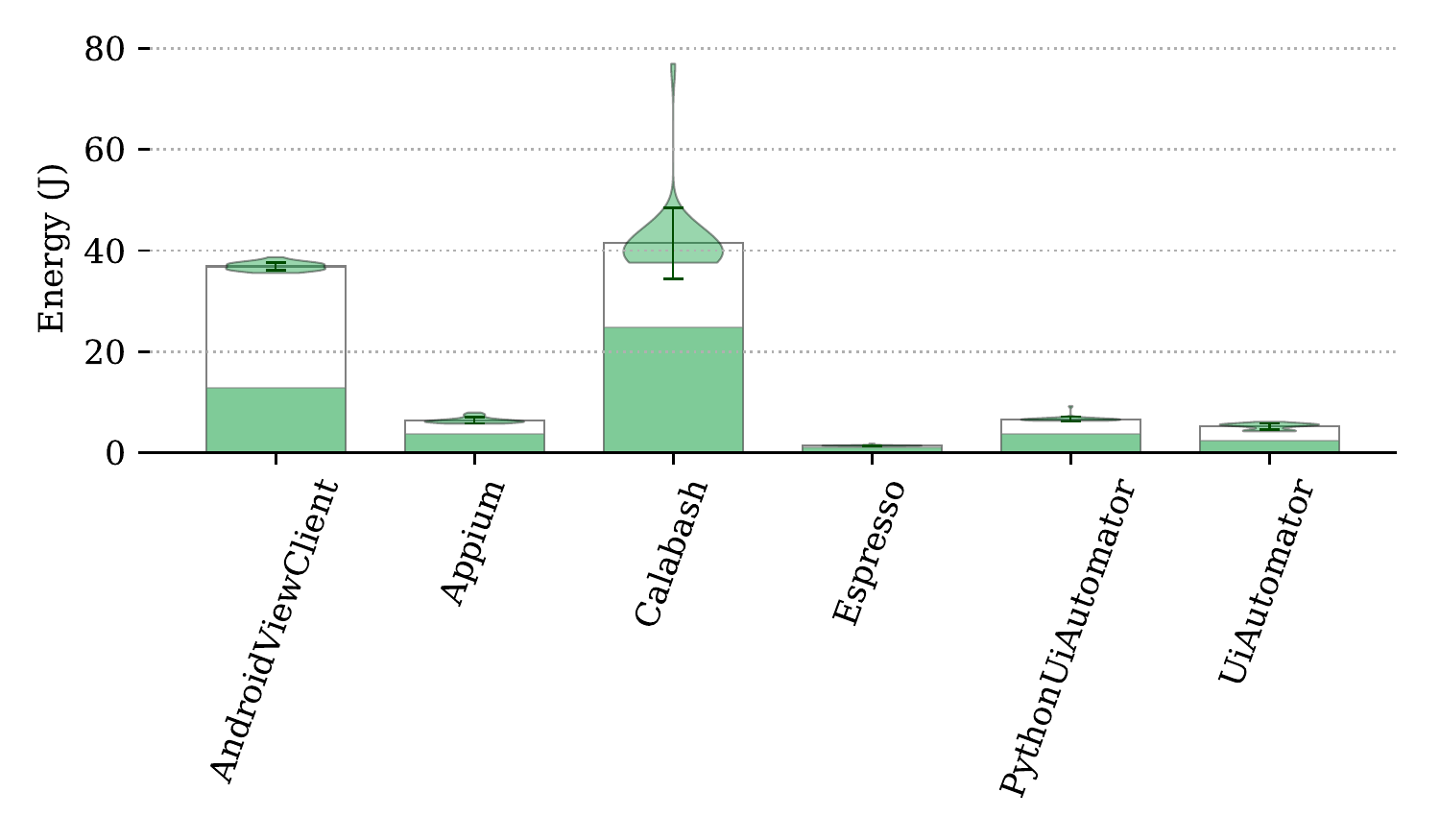}
  \caption{Violin plot of the results for energy consumption of \emph{Find by description}.}
  \label{fig:results_find_by_description}

\end{figure}

Results for \emph{Find by description} are presented in
Table~\ref{table_description_find_by_description} and
Figure~\ref{fig:results_find_by_description}. \emph{Find by description} and
\emph{Find by id} are very similar regarding usage and implementation, which
is confirmed by results. \emph{Espresso} is the best framework regardless of
idle cost ($\bar{x} = 1.37\textup{J}$ and $\bar{x}' = 0.97\textup{J}$).
\emph{Android View Client} and \emph{Calabash} are distinctly inefficient. All
other frameworks show reasonable energy footprints, except for \emph{Robotium}
and \emph{Monkeyrunner}, which were not included since \emph{Find by
description} is not supported.

\begin{table}
  \caption{Descriptive statistics of \emph{Find by description} interaction.}
  \label{table_description_find_by_description}
  \resizebox{1.0\linewidth}{!}{%
  \begin{tabular}{lrrrrrr}
\hline
                   &   $\bar{x}$ (J) &   $s$ &   $\Delta t$ (s) &   $\bar{{x}}'$ (J) &   Sg (mJ) &   Rank \\
\hline
 AndroidViewClient &           36.85 &  0.78 &           127.45 &              12.77 &     45.59 &      5 \\
 Appium            &            6.41 &  0.58 &            13.93 &               3.77 &     13.48 &      4 \\
 Calabash          &           41.41 &  7.02 &            88.20 &              24.75 &     88.38 &      6 \\
 \textbf{Espresso} &            1.37 &  0.10 &             2.10 &               0.97 &      3.46 &      1 \\
 PythonUiAutomator &            6.62 &  0.49 &            15.10 &               3.76 &     13.44 &      3 \\
 UiAutomator       &            5.13 &  0.61 &            14.47 &               2.40 &      8.57 &      2 \\
\hline
\end{tabular}

  }
\end{table}

\subsection{\emph{Find by content}}

\begin{figure}
    \centering
    \includegraphics[width=0.99\linewidth]{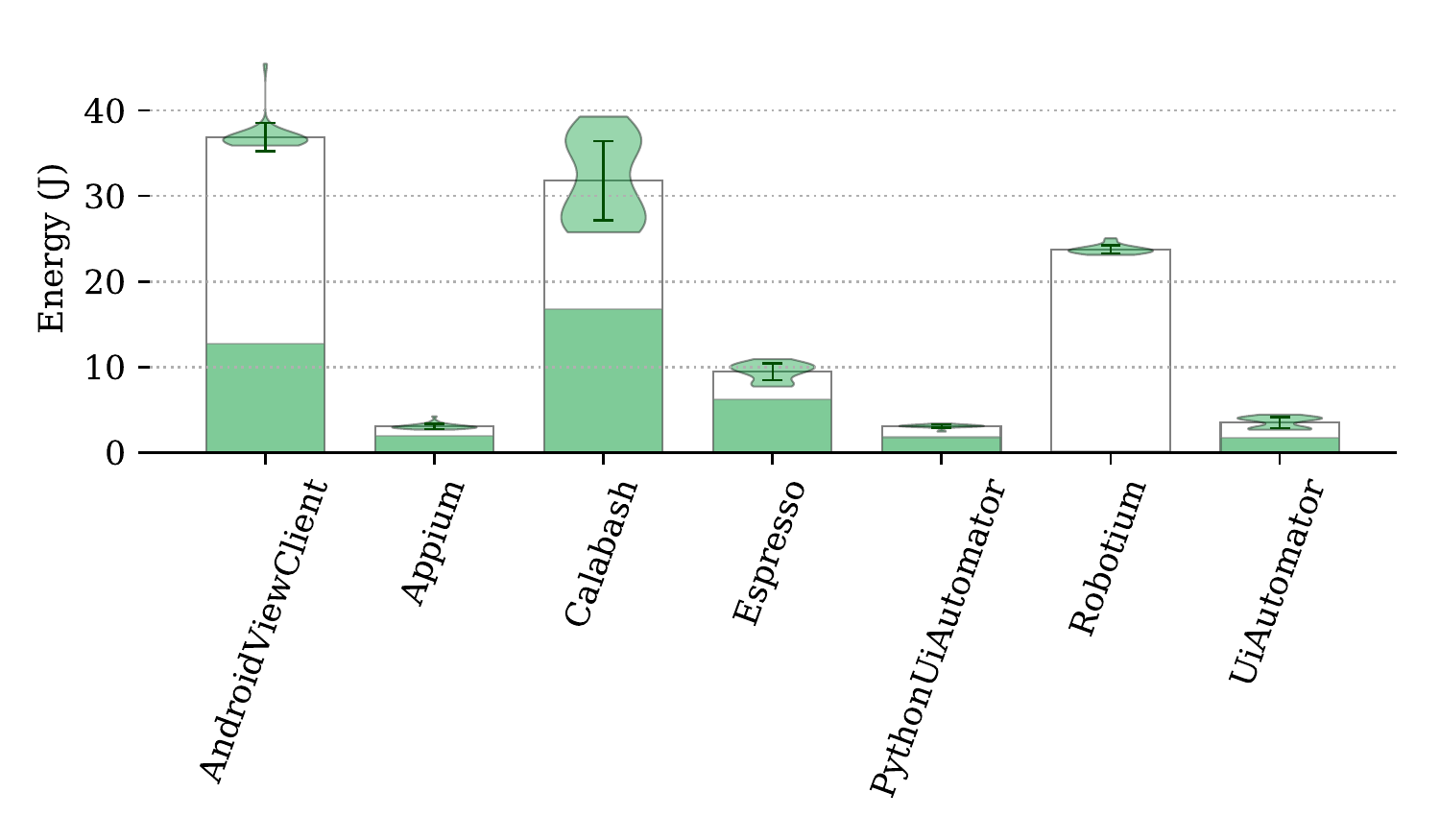}
  \caption{Violin plot of the results for energy consumption of \emph{Find by content}.}
  \label{fig:results_find_by_content}

\end{figure}

Results for \emph{Find by content} are presented in
Table~\ref{table_description_find_by_content} and
Figure~\ref{fig:results_find_by_content}. After removing idle cost,
\emph{Robotium} is the framework with best results ($\bar{x}' =
0.14\textup{J}$). However, in resemblance to \emph{Find by id}, Robotium is very
inefficient when idle cost is not factored out ($\bar{x} = 23.74\textup{J}$). In
this case, \emph{Appium} is the most efficient framework ($\bar{x} = 3.07\textup{J}$).

Unlike with \emph{Find by id} and \emph{Find by description}, \emph{Espresso}
did not yield good results in this case ($\bar{x} = 9.43\textup{J}$ and
$\bar{x}' = 6.19\textup{J}$). This is explained by the fact that
\emph{Espresso} runs natively on the DUT. Thus, finding a UI component by
content requires extra processing: the DUT has to search for a pattern in all
components' text content. Since remote script-based frameworks, such as \emph{Appium},
can do such task using the controller workstation, they can be more
energy efficient from the DUT's perspective. For the same reason, \emph{Find by
content} has consistently higher energy usage than the other helper methods.

\begin{table}
  \caption{Descriptive statistics of \emph{Find by content} interaction.}
  \label{table_description_find_by_content}
  \resizebox{1.0\linewidth}{!}{%
  \begin{tabular}{lrrrrrr}
\hline
                   &   $\bar{x}$ (J) &   $s$ &   $\Delta t$ (s) &   $\bar{{x}}'$ (J) &   Sg (mJ) &   Rank \\
\hline
 AndroidViewClient &           36.89 &  1.65 &           127.62 &              12.77 &    106.43 &      6 \\
 Appium            &            3.07 &  0.31 &             6.07 &               1.92 &     16.02 &      4 \\
 Calabash          &           31.77 &  4.64 &            79.63 &              16.72 &    139.35 &      7 \\
 Espresso          &            9.43 &  0.99 &            17.13 &               6.19 &     51.58 &      5 \\
 PythonUiAutomator &            3.10 &  0.19 &             6.90 &               1.79 &     14.93 &      3 \\
 \textbf{Robotium} &           23.74 &  0.48 &           124.90 &               0.14 &      1.15 &      1 \\
 UiAutomator       &            3.50 &  0.62 &             9.40 &               1.72 &     14.37 &      2 \\
\hline
\end{tabular}
}
\end{table}

\subsection{Statistical significance}

As expected from previous work and corroborated with the violin plots, our
measurements follow a normal distribution -- also confirmed with the
Shapiro-Wilk test. Thus, we assess the statistical significance of the mean
difference of energy consumption between frameworks using the parametric
Welch's t-test as used in previous work~\cite{cruz2017performance}. We apply
the Benjamini-Hochberg procedure by correcting p-values with the number of
times a given sample is used in different tests.

All but a few tests (2 out 105) resulted in a small $p$-value, below the
significance level $\alpha=0.05$. For those pairs where there was no
statistical significance, we could not find any meaningful finding. Given the
myriad number of tests performed, results are not presented. Violin plots
corroborate statistical significance by presenting very distinct distributions
among all different frameworks. For further details, all results and data are
publicly available\footnote{Project's \emph{Github} repository:
\url{https://github.com/luiscruz/physalia-automators} visited on \today{}.}.

\subsection{Threats to validity}

\paragraph*{Construct validity}

Frameworks rely on different approaches to collect information about the UI
components that are visible on the screen. The app used in the experiments has
a UI that remains unchanged upon user interactions. In a real scenario, however,
the UI typically reacts to user interactions. Frameworks that have an
inefficient way of updating their UI model of components visible in the screen,
may entail a high overhead on energy consumption. However, as manually
triggering this update is not supported in most frameworks, it was unfeasible
to include it in our study.

In addition, the overheads are calculated based on the results collected from the
human interaction from two participants. Although results showed a small
variance between different participants, the energy consumption may vary with
other humans. Nevertheless, differences are not expected to be significant, and
results still apply.

Moreover, energy consumption for a single interaction is inferred by the
total consumption of a sequence of interactions. Potential \emph{tail energy
consumptions}\footnote{\emph{Tail energy} is the energy
spent during initialization or closure of a resource.} of a single interaction
are not being measured. This is mitigated by running multiple times the same
interaction.


\paragraph*{Internal validity} The Android OS is continuously running parallel
tasks that affect energy consumption. For that reason,
system settings were customized as described in Section~\ref{sec:methodology}
(e.g., disabled automated brightness and notifications). Also, each
experiment is executed $30$ times to ensure statistical significance as
recommended in related work~\cite{linares2014mining}.

UI interactions typically trigger internal tasks in the mobile application
running in foreground. The mobile application used in experiments was developed
to prevent any side-effects to UI events. To ensure that scripts are
interacting with the device as expected, the application was set to a mode that
is not affected by user interaction. Thus, the behavior is equal across
different UI automation frameworks and experiments only measure their energy
consumption.

Finally, our experiments use a WiFi-configured ADB instead of a USB connection.
This is a requirement from remote script-based frameworks. We did not measure
the energy consumption entailed from using a USB-configured ADB. Nevertheless,
we do not expect results to differ since the WiFi connection is only used
before and after the measurements.

\paragraph*{External validity} Energy consumption results vary upon different
versions of Android OS, different device models, and different framework
version. However, unless major changes are released, results are not expected
to significantly deviate from the reported ones. Note that testing different
devices requires disassembling them and making them useless for other
purposes (that is to say that empirical studies as the one conducted by us are
expensive), which can be economically unfeasible. Regardless, all the source code
used in experiments will be released as Open Source to foster reproducibility.
\section{Discussion}
By answering the research questions, in this section we discuss our findings
from the empirical evaluation, as well as outline their practical implications.

\rquestion{1}{Does the energy consumption overhead created by UI
automation frameworks affect the results of the energy efficiency of mobile
applications?}

Yes, results show that interactions can have a tremendous overhead on energy
consumption when an inefficient UI automation framework is used.

According to previous work, executing a real app during $100\textup{s}$ yields
an energy consumption of $58\textup{J}$, on average~\cite{li2014empirical}.
Considering our results, executing a single interaction such as \emph{Drag and
Drop} can increase energy consumption in $1.7\textup{J}$ (overhead of $3\%$ in
this case). However, given that mobile apps are very reactive to user
input~\cite{joorabchi2013real}, in $100$ seconds of execution, more
interactions are expected to affect energy. Although a fair comparison must
control for different devices and OS versions, this order of magnitude implies
that overheads are not negligible. Thus, choosing an efficient UI automation
framework is quintessential for energy tests.


Since all frameworks produce the same effect in the UI, the overhead of energy
consumption is created by implementation decisions of the framework and not by
the interaction itself. The main goal of a UI testing framework is to mimic
realistic usage scenarios, but interactions with such overhead can be
considered unrealistic.

One practical implication of the results in this work is to drive a change in the
mindset of tool developers, bringing awareness of the energy consumption of
their frameworks. Thus, we expect future releases of UI automation frameworks
to become more energy efficient.

\emph{AndroidViewClient} and \emph{Calabash} consistently showed poor energy
efficiency among all interactions. Despite providing a useful and complete
toolset for mobile software developers, they should be used with prudence while
testing the energy consumption of an app that heavily relies on user
interactions. Work of Carette A. \emph{et al.} (2017) [11] was affected by a
poor choice of framework: the authors used \emph{Calabash} to mimic between
$136$ and $325$ user interactions in experiments that, in total, consume
roughly $350$J. Considering our results, a single \emph{tap} with Calabash is
equivalent to $0.45$J -- it means that at least $60$J ($17$\%) were spent by
the UI framework. The same interactions with \emph{Espresso} would have been
reduced to $12$J ($3$\%). The impact increases when considering other
interactions. Our work shows that results would be different if the overhead of
the framework had been factored out. On the contrary, \emph{Calabash} was also
used in other work [12] but its impact can be considered insignificant since
experiments did not require much interaction and the main source of energy
consumption came from Web page loads. Note, however, that the measurement setup
is different and results from related work are not directly comparable. We plan
to address this analysis in future work. In any case, we consider that using a
more energy efficient framework could corroborate the evidence or find new --
even contradictory -- conclusions.

\rquestion{2}{What is the most suitable framework to profile energy consumption?}

\begin{figure*}
    \centering
    \includegraphics[width=0.8\linewidth]{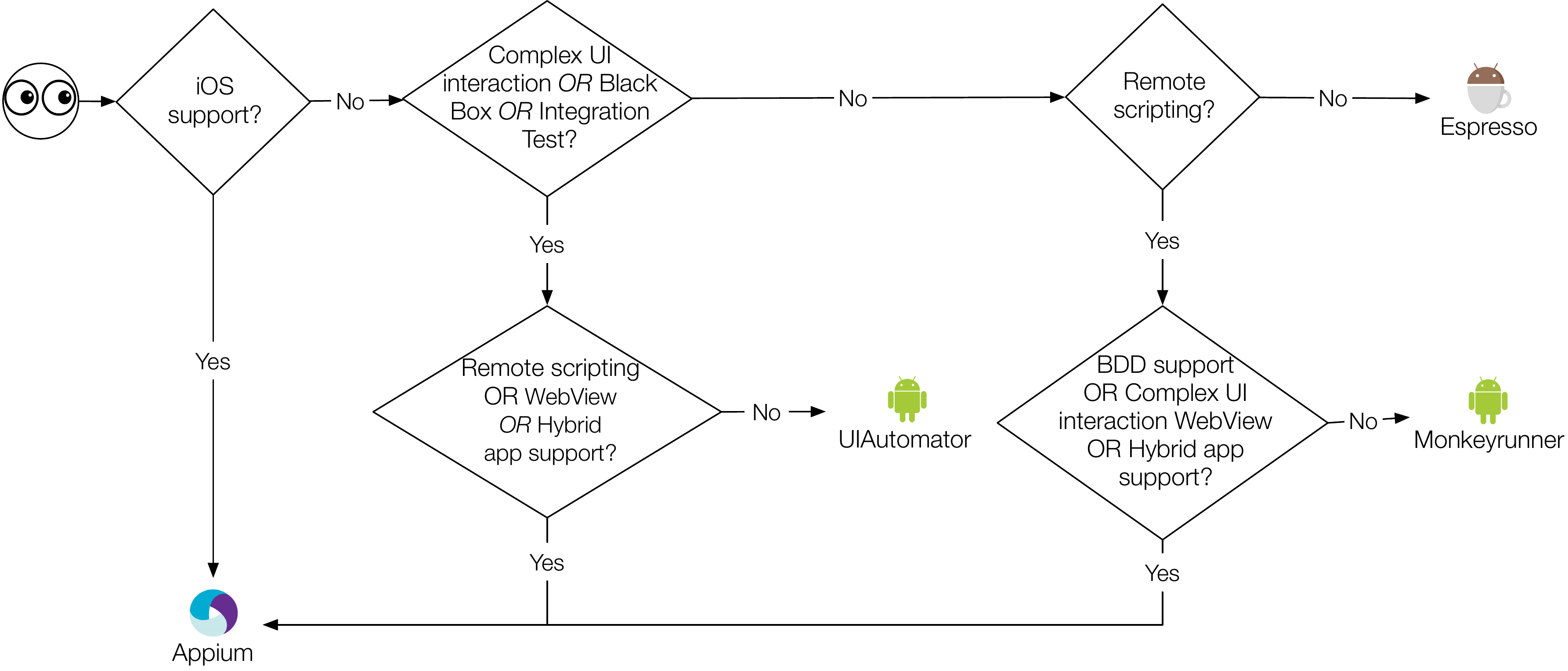}
  \caption{Selecting the most suitable framework for energy measurements.}
  \label{fig:decision_diagram}
\end{figure*}

Choosing the right framework for a project can be challenging: there is no
\textit{one solution fits all}. Based on our observations,
Figure~\ref{fig:decision_diagram} depicts a decision tree to help software
developers making an educated guess about the most suited and energy efficient
framework, given the idiosyncrasies of an app (that may restrict the usage of a
framework). For example, if the project to be tested requires \emph{WebView} support,
one should use \emph{Appium} rather than the other frameworks. \emph{Robotium}
is also an option if the app requires \emph{Taps} or \emph{Input Text} only,
and neither iOS support nor remote scripting is required.

Remote script-based frameworks allow developers to easily create automation
scripts. The script can be iteratively created using a console while
interactions take effect on the phone in real time. From our experience while
doing this work, remote script-based frameworks are easier to use and set up
(i.e., gradual learning curve). This is one of the reasons many frameworks
decided to use scripting languages (e.g., Python and Ruby) instead of the
official languages for Android, \emph{Java} or \emph{Kotlin}. Notwithstanding,
remote script-based frameworks require an active connection with the phone
during measurements, which leads to higher energy consumption (as is confirmed
by results). Each step of the interaction requires communication with the DUT;
hence, the communication logic unavoidably increases the energy consumption. On
the contrary, other frameworks can transfer the interaction script in advance
to the mobile phone and run it natively on the phone, which is more energy
efficient.

There are, however, two scenarios where remote script-based frameworks exhibit
the best results: \emph{Back Button} with \emph{Monkeyrunner} (see
Table~\ref{table_description_back_button}), and \emph{Find By Content} with
\emph{Appium} (see Table~\ref{table_description_find_by_content}). This is an
interesting finding as it shows that remote script-based frameworks can also be
developed in an energy efficient way. As such, this evidence shows that there
is room for energy optimization in the other frameworks.

In addition, USB communication is out of question for remote script-based
frameworks since it affects the reliability of measurements. Frameworks that do
not support remote scripting can be used with USB connection if unplugged
during measurements (using tools such as \emph{Monsoon Power Monitor}).

Among remote script-based frameworks, \emph{Monkeyrunner} is the most energy efficient
framework. The only problem is that it does not support many of the studied
interactions. These results show that if energy consumption turns into a priority, it
is possible to make complex frameworks such as \emph{Appium} more energy
efficient.

There are a number of other fine-grained requirements that developers need to
consider when choosing a UI framework. A more thorough decision ought to
consider other factors, such as existing infrastructure, development process,
and learning curve. Nevertheless, we argue that decision tree of
Figure~\ref{fig:decision_diagram} provides an approximate insight even though
it does not take all factors into account.


\rquestion{3}{Are there any best practices when it comes to creating automated
scripts for energy efficiency tests?}

One thing that stands out is the fact that looking up one UI component is
expensive. This task is exclusively required for automation and does not
reflect any real-world interaction. Taking the example of \emph{Espresso}: a
single \emph{Tap} consumes $0.09\textup{J} $, while using content to look up a component
consumes $0.05\textup{J}$. Since a common \emph{Tap} interaction requires
looking a component up, 36\% of energy spent is on that task.

Looking up UI components is energy greedy because the framework needs to
process the UI hierarchy find a component that matches a given id, description,
or content. Since the app we use has a very simple UI hierarchy, the energy
consumption is likely to be higher in real apps. Hence, using \textbf{lookup methods
should be avoided whenever possible}. A naive solution could be using the pixel
position of UI components instead of identifiers. Pixel positions could be
collected using a recorder. However, this is a bad practice since it brings
majors maintainability issues across different releases and device models. For
that reason, state-of-the-art UI recorders used by Android developers, such as
\emph{Robotium Recorder}, yield scripts based on UI identifiers. As an
alternative, we recommend caching the results of lookup calls whenever possible.

In addition, \textbf{lookup methods \emph{Find by Id} and \emph{Find by Description}
should be preferred to \emph{Find by Content}}. Results consistently show worse
energy efficiency when using \emph{Find by Content}. In \emph{Espresso}, this
difference gives an increase in energy consumption from $1.4\textup{J}$ to
$9.4\textup{J}$ (overhead of $600\%$).

\vspace{-1.5em}
\section{Conclusion}

In this paper, we analyze eight popular UI automation frameworks for mobile
apps with respect to their energy footprint. UI interactions have distinct
energy consumptions depending on the framework. Our results show that the
energy consumption of UI automation frameworks \textbf{should be factored out to avoid
affecting results of energy tests}. As an example, we have observed the overhead
of the \emph{Drag and Drop} interaction to go up to $2200\%$. Thus,
practitioners and researchers should opt for energy efficient frameworks.
Alternatively, the energy entailed by automated interactions must be factored
out from measurements.

\textbf{\emph{Espresso} is observed to be the most energy efficient framework}.
Nevertheless, it has requirements that may not apply to all projects:
1) requires access to the source code, 2) does not support
complex interactions such as \emph{Drag and Drop} and \emph{Pinch and Spread}, 3)
is not compatible with WebViews, 4) is OS dependent, and 5) is not remote script-based.
Hence, there are situations where \emph{UIAutomator}, \emph{Monkeyrunner}, and
\emph{Appium} are also worth considering. \textbf{For a more general purpose context,
\emph{Appium} follows as being the best candidate}. Thus,
we propose a decision tree (See Figure~\ref{fig:decision_diagram}) to help in
the decision-making process.

Furthermore, we have also noticed the following in our experiments. Helper
methods to find components in the interface are necessary when building energy tests,
but should be minimized to prevent affecting energy results. In particular, \textbf{lookup
methods based on the content of the UI component need to be avoided}. They consistently
yield poor energy efficiency when
compared to lookups based on id (e.g., in \emph{Espresso} it creates an overhead
of $600\%$).

This work paves the way for the inclusion of energy tests in the development
stack of apps. It brings awareness to the energy footprint of tools used for
energy test instrumentation, affecting both academic and industrial use cases.
It remains to future work to design a catalog of energy-aware testing
patterns\footnote{For instance, augmenting/improve the following list of testing
patterns \url{http://wiki.c2.com/?TestingPatterns}}.
\vspace{-1.0em}
\section*{Acknowledgments}
\vspace{-0.3em}
{\small
\noindent This work was partially funded by FCT - Fundação para a Ciência e a
Tecnologia with reference UID/EEA/50014/2019, the GreenLab
Project (ref. POCI-01-0145-FEDER-016718), the FaultLocker Project (ref.
PTDC/CCI-COM/29300/2017), and the Delft Data Science (DDS) project.}
\balance
\vspace{-1.2em}
\bibliographystyle{IEEEtran}
\bibliography{bibliography}%
\begin{IEEEbiography}%
[{\includegraphics[width=1in,height=1.25in,clip,keepaspectratio]{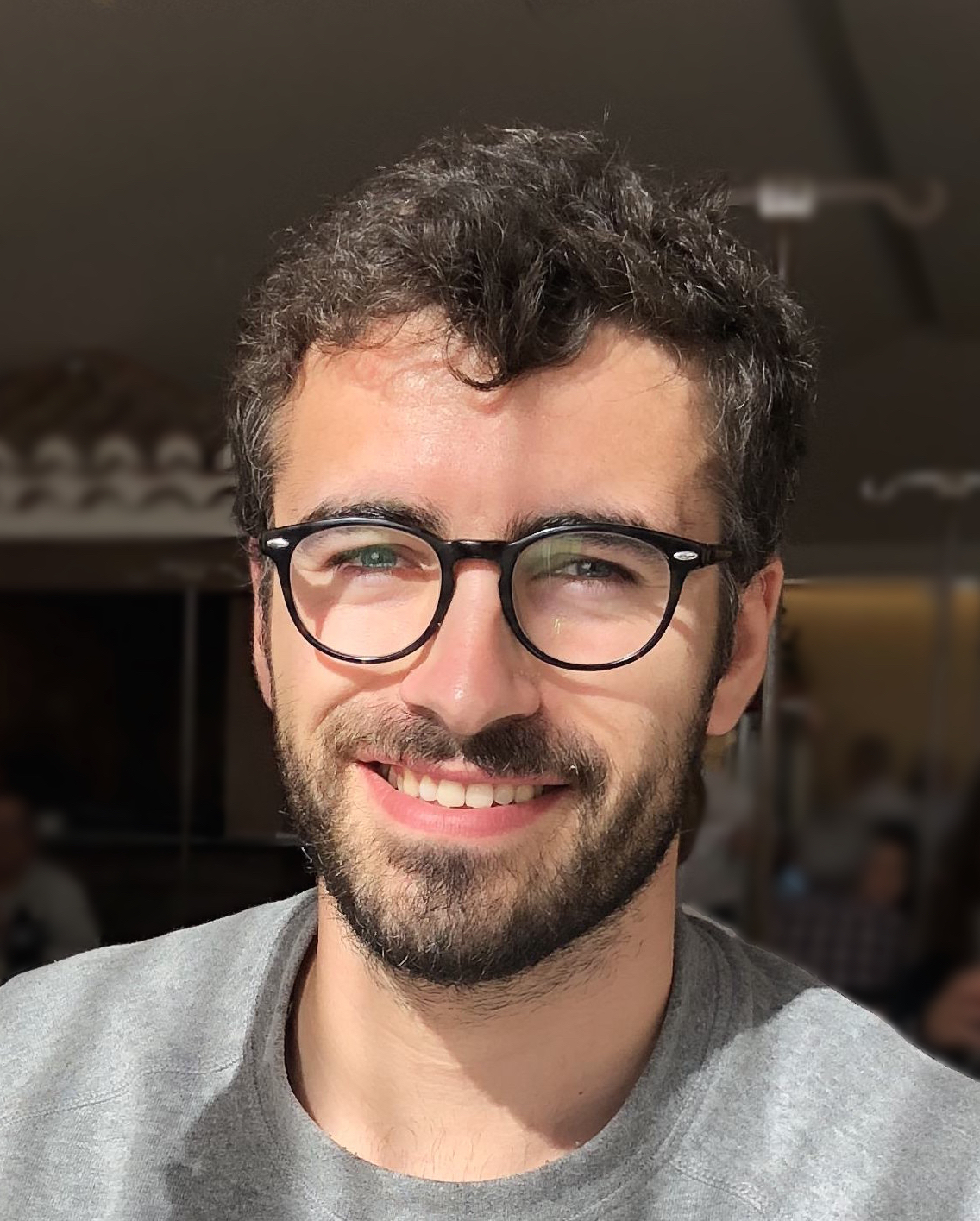}}]%
{Lu\'{i}s Cruz}%
 holds a Ph.D. in Computer Science from the University of Porto, Portugal. He is currently a postdoc researcher
in Delft University of Technology where most of his research is carried out. His main research fields are Mobile Software Engineering,
Green Computing, and Mining Software Repositories. His work aims at improving mobile development processes
concerning the energy consumption of mobile applications.

\end{IEEEbiography}%
\begin{IEEEbiography}%
[{\includegraphics[width=1in,height=1.25in,clip,keepaspectratio]{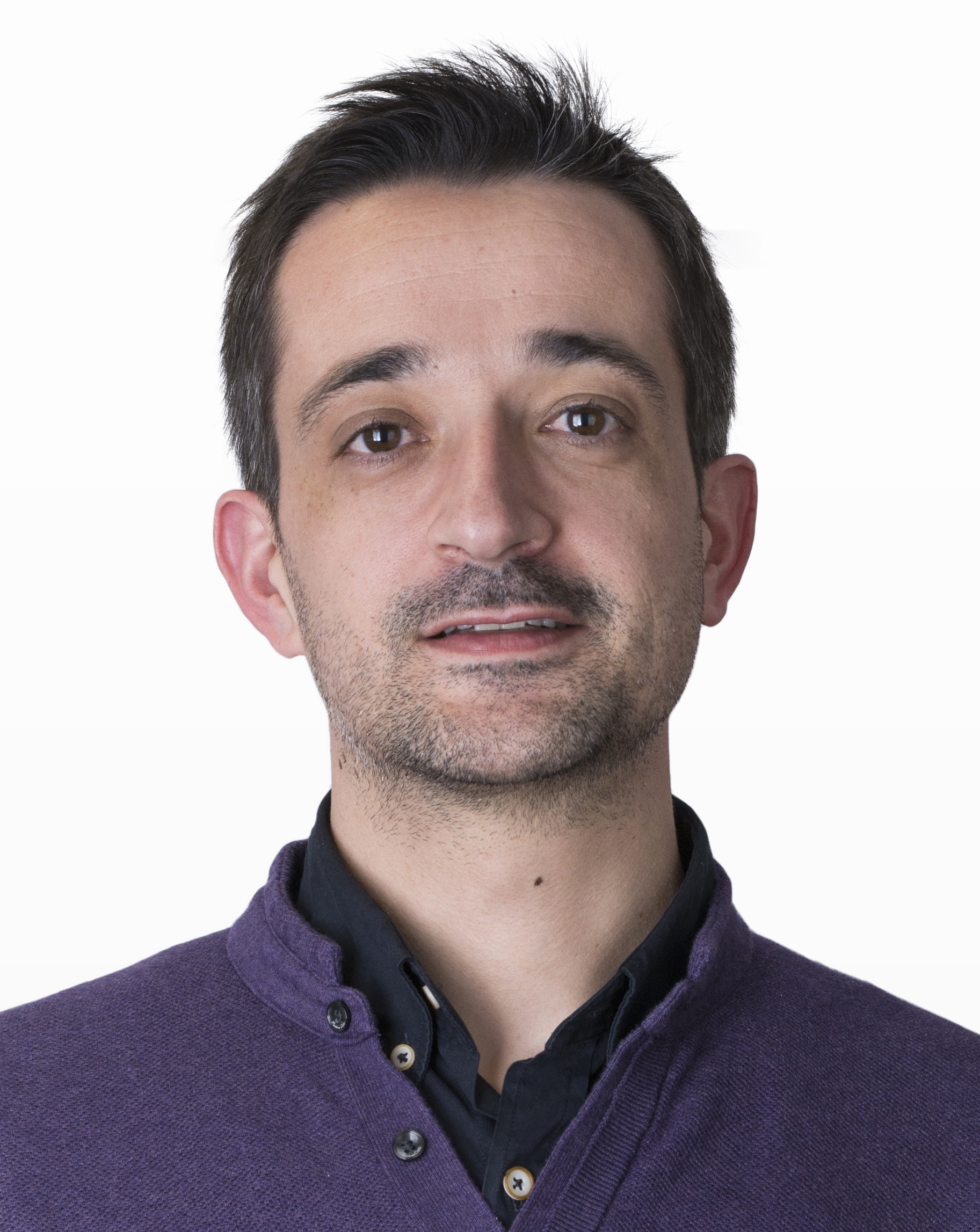}}]%
{Rui Abreu}%
 holds a Ph.D. in Computer Science - Software Engineering from
the Delft University of Technology, The Netherlands, and a M.Sc. in Computer
and Systems Engineering from the University of Minho, Portugal. His research
revolves around software quality, with emphasis in automating the testing and
debugging phases of the software development life-cycle as well as
self-adaptation. Dr. Abreu has extensive expertise in both static and dynamic
analysis algorithms for improving software quality. He is the recipient of 6
Best Paper Awards, and his work has attracted considerable attention. Before
joined IST, ULisbon as an Associate Professor and INESC-ID as a Senior
Researcher, he was a member of the Model-Based Reasoning group at PARC’s System
and Sciences Laboratory and an Assistant Professor at the University of Porto.
He has co-founded DashDash in January 2017, a platform to create web apps using
only spreadsheet skills. The company has secured \$9M in Series A funding in May
2018.%
\end{IEEEbiography}%
\end{document}